\documentclass[letterpaper,11pt,twoside]{article}
\usepackage{amsmath,amssymb,amsthm}
\usepackage{epic,eepic}

\newcommand{\xd}{\mathrm{d}}
\newcommand{\tens}{\otimes}
\newcommand{\defeq}{:=}
\newcommand{\Z}{\mathbb{Z}}
\newcommand{\C}{\mathbb{C}}
\newcommand{\N}{\mathbb{N}}
\DeclareMathOperator{\cou}{\epsilon}
\DeclareMathOperator{\cop}{\Delta}
\newcommand{\one}{\mathbf{1}}
\DeclareMathOperator{\id}{id}
\newtheorem{prop}{Proposition}
\newtheorem{thm}[prop]{Theorem}
\newtheorem{cor}[prop]{Corollary}
\newtheorem{lem}[prop]{Lemma}
\newtheorem{dfn}[prop]{Definition}
\newcommand{\Rop}{R}
\newcommand{\Qop}{Q}
\newcommand{\Dop}{\Lambda}
\newcommand{\Ropm}{\hat{R}}
\newcommand{\Qopm}{\hat{Q}}
\newcommand{\Dopm}{\hat{\Lambda}}
\newcommand{\Gcompl}{G}
\newcommand{\Gfeyn}{G_F}
\newcommand{\Gconn}{G_c}
\newcommand{\Gopi}{G_{\text{1PI}}}
\newcommand{\Gopim}{\hat{G}_{\text{1PI}}}

\newcommand{\compl}{\rho}
\newcommand{\conn}{\sigma}
\newcommand{\connm}{\hat{\sigma}}
\newcommand{\opi}{\tau}
\newcommand{\opim}{\hat{\tau}}
\newcommand{\SV}{\mathsf{S}(V)}
\newcommand{\complt}{\rho_T}
\newcommand{\connt}{\sigma_T}
\newcommand{\opit}{\tau_T}

\allowdisplaybreaks

\title{\textbf{Combinatorics of $n$-point functions via Hopf algebra
 in quantum field theory}}
\author{\^Angela Mestre\footnote{email: mestre@matmor.unam.mx},
 Robert Oeckl\footnote{email: robert@matmor.unam.mx}\\ \\
Instituto de Matem\'aticas, UNAM Campus Morelia,\\
C.~P.~58190, Morelia, Michoac\'an, Mexico}
\date{24 May 2005\\ 27 January 2006 (v2)}

\begin{document}

\maketitle

\begin{abstract}
We use a coproduct on the time-ordered algebra of field operators to
derive simple relations between complete, connected and 1-particle
irreducible $n$-point functions. Compared to traditional functional
methods our approach is much more intrinsic and leads to efficient
algorithms suitable for concrete computations. It may also be used to
efficiently perform tree level computations.

\end{abstract}

\section{Introduction}

It may be said that time-ordered $n$-point functions are at the heart
of (perturbative) quantum field theory. These determine the S-matrix
that in turn allows to calculate experimentally observable scattering
cross sections. Besides the complete $n$-point functions which are
expectation values of time-ordered products of field operators,
a key role is played by connected and by 1-particle irreducible (1PI)
$n$-point functions. In particular, the latter play a prominent role in
the process of renormalization, as it is enough to renormalize 1PI
$n$-point functions. Furthermore, they are intimately related to the
effective action. The different classes of $n$-point functions
correspond directly to different sums over Feynman graphs, namely all
graphs (excluding vacuum graphs), connected graphs and 1PI graphs.

The relations between these different classes of $n$-point functions
(and thus sums over Feynman graphs) are traditionally expressed by using
functional methods. While having an undeniable elegance, a
disadvantage of these methods is the need for auxiliary sources and
associated generating functions. These, as well as functional
derivatives used in the process are often purely formal and have no
rigorous mathematical existence. Consequently, they are only
indirectly related to concrete calculations of $n$-point functions one
may wish to perform.

In the present article we describe the relation between these classes
of $n$-point functions directly on the level of the algebra of
time-ordered field operators. The key structure we will make use of
is the natural \emph{coproduct} on this algebra, making it into a
\emph{Hopf algebra}. Ensembles of time-ordered $n$-point functions are
simply linear forms on this algebra. We show that the convolution
product (induced by the coproduct) provides an extremely concise and
elegant way of relating complete and connected $n$-point functions.
Indeed, this relation is simply given by the convolution exponential (or,
conversely, the logarithm).

Our second (and perhaps main) result concerns the relation of
connected and 1PI $n$-point functions. As is well known, the former
are expressible in terms of the latter as the sum over all tree graphs
with 1PI vertices. We present a simple recursion formula using the
coproduct, which generates exactly all tree
diagrams. Moreover, the result takes an algebraic form which can be
directly evaluated on 1PI functions so as to yield the connected
functions.
We proceed to derive an alternative recursion formula which relates
directly components of connected $n$-point functions ordered by
vertex number. Moreover, these same formulas can be alternatively
applied to calculate the tree level contribution to the connected
$n$-point functions, using the interaction terms of the Lagrangian.

A key feature of our results is their close
relation to algorithmic descriptions of the computations involved.
Indeed, it is easy to read off from our recursion relations not only
algorithms to perform the computations, but even data structures
relevant for an implementation. Our algorithm for
generating trees also seems to be particular efficient as
it allows to impose a lower bound on the number of legs per vertex
from the outset.

The present article may be seen as part of a programme, laid out in
\cite{BFFO:twist} and rooted in \cite{Fau:wick,Bro:qfieldalg,Oe:bqft},
with the aim of formulating and understanding the 
key structures of quantum field theory and their
combinatorics in terms of the Hopf algebra of field operators.
The focus of \cite{BFFO:twist} were the different products of field
operators (normal product, canonical product and time-ordered
product), uncovering their relation through Drinfeld twists via
certain 2-cocycles in Hopf algebra cohomology.
At the same time the relation between products and
associated $n$-point functions was elucidated, again using Hopf
algebra cohomology. The present article complements this by
investigating with the same means
the relation between different classes of $n$-point
functions that correspond to different classes of Feynman diagrams.

While Hopf algebras and coproducts have a long history in
combinatorics, their use in combinatorial problems in quantum field
theory is rather recent. The probably first instance was Kreimer's
Hopf algebraic explanation of the Bogoliubov formula of
renormalization and of Zimmermann's solution \cite{Kre:hopfpert},
subsequently developed together with Connes \cite{CoKr:renriem1}.
We caution the reader, however, that the Hopf
algebras used by
Kreimer and Connes, while also being related to Feynman graphs, are
quite distinct from the Hopf algebra of field operators used
here.

Section~\ref{sec:basics} starts with recalling the different classes
of $n$-point functions and their relation to Feynman graphs. Then, the
basic algebraic formalism used in this article is introduced, in
particular the coproduct.
Section~\ref{sec:complconn} deals with the relation between complete
and connected $n$-point function, Section~\ref{sec:conn1pi} with that
between connected and 1PI $n$-point functions. An alternative
recursion formula for the latter is derived in
Section~\ref{sec:recrel}. In Section~\ref{sec:discuss} various
generalizations (e.g.\ tree level calculations) and related issues are
discussed. Some Conclusions are offered in
Section~\ref{sec:concl}. The appendix shows all tree graphs with up to
$7$ vertices with weight factors computed according to
Section~\ref{sec:conn1pi}.

No knowledge of Hopf algebras is required to read this article.

\section{Basic definitions}
\label{sec:basics}

We shall be concerned in the following with a generic perturbative
quantum field theory. We denote the basic field operators by
$\phi(x)$, where $x$ represents a label that completely determines the
operator. In a position representation $x$ would specify a point in
Minkowski space, possibly together with internal indices. While our
notation suggests a field theory with a single scalar field, this is
just a convenience. Although our results are general, we limit
ourselves in the following exposition to a purely bosonic theory for
simplicity. We return in Section~\ref{sec:fermions} to a discussion of
the general case, including fermionic fields.

\subsection{Feynman graphs and $n$-point functions}
\label{sec:review}

We review here essentials about (classes of) Feynman graphs and
$n$-point functions. For more information on these and the standard
functional approach used to manipulate them we refer the reader to
standard text books such as \cite{ItZu:qft}.

\begin{dfn}
A \emph{graph} is a finite collection of \emph{vertices} and
\emph{edges} (also called \emph{legs}), such that any
end of an edge may be connected to a vertex.
Edges that are connected to vertices at both ends are
called \emph{internal}, while edges with at least one free end are
called \emph{external}.
The \emph{valence} of a vertex is the number of ends of edges
connected to the vertex. A \emph{tree} (graph) is a connected graph
that has no cycles.
\end{dfn}

Feynman graphs are graphs that carry certain labels on vertices and
edges. The former correspond to the interaction terms of the
Lagrangian while the latter may correspond to momenta and internal
indices (usually indicated by different styles of lines, e.g.,
straight for fermions, wiggly for bosons etc.).
We shall
only consider labels attached to (open ends) of external legs, assuming
internal labels to be summed or integrated over.
\begin{dfn}
A \emph{labeled} graph is a graph whose (ends of) external legs are
labeled by field operator labels.
\end{dfn}
In the following we shall consider only such labeled graphs, i.e.,
from now on \emph{graph} really means \emph{labeled graph}. A Feynman
graph has a (usually complex) value as a function of the labels on the
external legs.

We denote by
$\Gcompl^{(n)}(x_1,\dots,x_n)$ the \emph{complete} 
$n$-point function. This is the vacuum expectation value of
the time-ordered product of $n$ field operators, i.e.,
\[
 \Gcompl^{(n)}(x_1,\dots,x_n)
 =\langle 0 | T \phi(x_1)\cdots \phi(x_n) | 0\rangle .
\]
In terms of Feynman graphs it is the sum of the values of all
graphs with external legs labeled by $x_1,\dots,x_n$. Let us
denote by $\Gamma^n$ the set of all Feynman graphs of the given
theory\footnote{Whether one considers the bare or renormalized
(including counter terms) theory does not matter to us here. Of course,
in the former case quantities might be infinite and manipulations
therefore formal.} with $n$ legs (with vacuum graphs, i.e., graphs
containing pieces not connected to any external leg already
excluded). For a given graph
$\gamma\in\Gamma^n$ we denote by $\gamma(x_1,\dots,x_n)$ its value for
the given labelings. Then,
\begin{equation}
 \Gcompl^{(n)}(x_1,\dots,x_n)=\sum_{\gamma\in\Gamma^n}
 \gamma(x_1,\dots,x_n) .
\label{eq:npfeyn}
\end{equation}

Of particular importance is the Feynman propagator $\Gfeyn(x,y)$
which is the value of the graph that consists of an edge only, its two
ends labeled by $x$ and $y$ respectively. Note that the Feynman
propagator is symmetric in its arguments $\Gfeyn(x,y)=\Gfeyn(y,x)$ as
suggested by the corresponding symmetry of the graph.
We
also define its inverse $\Gfeyn^{-1}$ which is determined by the
equation\footnote{Here as in the following we use a notation
that suggests merely an integration over space-time. However,
appropriate summations over internal indices are also implied, but not
written explicitly.}
\begin{equation}
\int \xd y\, \Gfeyn(x,y)\Gfeyn^{-1}(y,z)=\delta(x,z) .
\label{eq:invfey}
\end{equation}

\begin{figure}
\setlength{\unitlength}{0.00083333in}
\begingroup\makeatletter\ifx\SetFigFont\undefined%
\gdef\SetFigFont#1#2#3#4#5{%
  \reset@font\fontsize{#1}{#2pt}%
  \fontfamily{#3}\fontseries{#4}\fontshape{#5}%
  \selectfont}%
\fi\endgroup%
{\renewcommand{\dashlinestretch}{30}
\begin{picture}(5777,1140)(0,-10)
\path(19,1006)(919,106)
\path(12,114)(912,1014)
\path(1159,1013)(2059,113)
\path(1152,121)(2052,1021)
\path(3702,121)(4617,121)
\path(3702,121)(4617,121)
\path(3687,1020)(4602,1020)
\path(3687,1020)(4602,1020)
\path(4865,1013)(5120,563)
\path(4865,1013)(5120,563)
\path(5765,1028)(5555,578)
\path(5765,1028)(5555,578)
\path(4857,106)(5517,541)
\path(4857,106)(5517,541)
\path(5757,121)(5067,571)
\path(5757,121)(5067,571)
\path(3319,106)(3319,1021)
\path(3319,106)(3319,1021)
\path(2419,106)(2419,1021)
\path(2419,106)(2419,1021)
\put(2052,504){\makebox(0,0)[lb]{{\SetFigFont{12}{14.4}{\rmdefault}{\mddefault}{\updefault}$+$}}}
\put(3499,504){\makebox(0,0)[lb]{{\SetFigFont{12}{14.4}{\rmdefault}{\mddefault}{\updefault}$+$}}}
\put(4647,504){\makebox(0,0)[lb]{{\SetFigFont{12}{14.4}{\rmdefault}{\mddefault}{\updefault}$+$}}}
\put(987,519){\makebox(0,0)[lb]{{\SetFigFont{12}{14.4}{\rmdefault}{\mddefault}{\updefault}$=$}}}
\put(4137,71){\makebox(0,0)[lb]{{\SetFigFont{12}{14.4}{\rmdefault}{\mddefault}{\updefault}c}}}
\put(4137,71){\makebox(0,0)[lb]{{\SetFigFont{12}{14.4}{\rmdefault}{\mddefault}{\updefault}c}}}
\put(4122,970){\makebox(0,0)[lb]{{\SetFigFont{12}{14.4}{\rmdefault}{\mddefault}{\updefault}c}}}
\put(4122,970){\makebox(0,0)[lb]{{\SetFigFont{12}{14.4}{\rmdefault}{\mddefault}{\updefault}c}}}
\put(461,563){\whiten\ellipse{210}{210}}
\put(461,563){\ellipse{210}{210}}
\put(1601,570){\whiten\ellipse{210}{210}}
\put(1601,570){\ellipse{210}{210}}
\put(4174,113){\whiten\ellipse{210}{210}}
\put(4174,113){\ellipse{210}{210}}
\put(4159,1012){\whiten\ellipse{210}{210}}
\put(4159,1012){\ellipse{210}{210}}
\put(5127,555){\whiten\ellipse{210}{210}}
\put(5127,555){\ellipse{210}{210}}
\put(5516,563){\whiten\ellipse{210}{210}}
\put(5516,563){\ellipse{210}{210}}
\put(3305,578){\whiten\ellipse{210}{210}}
\put(3305,578){\ellipse{210}{210}}
\put(2405,578){\whiten\ellipse{210}{210}}
\put(2405,578){\ellipse{210}{210}}
\put(1556,518){\makebox(0,0)[lb]{{\SetFigFont{12}{14.4}{\rmdefault}{\mddefault}{\updefault}c}}}
\put(4122,67){\makebox(0,0)[lb]{{\SetFigFont{12}{14.4}{\rmdefault}{\mddefault}{\updefault}c}}}
\put(4107,966){\makebox(0,0)[lb]{{\SetFigFont{12}{14.4}{\rmdefault}{\mddefault}{\updefault}c}}}
\put(5082,511){\makebox(0,0)[lb]{{\SetFigFont{12}{14.4}{\rmdefault}{\mddefault}{\updefault}c}}}
\put(5472,511){\makebox(0,0)[lb]{{\SetFigFont{12}{14.4}{\rmdefault}{\mddefault}{\updefault}c}}}
\put(3267,526){\makebox(0,0)[lb]{{\SetFigFont{12}{14.4}{\rmdefault}{\mddefault}{\updefault}c}}}
\put(2367,526){\makebox(0,0)[lb]{{\SetFigFont{12}{14.4}{\rmdefault}{\mddefault}{\updefault}c}}}
\end{picture}
}
\caption{Decomposition of the complete $4$-point function in terms of
  connected functions.}
\label{fig:compl4}
\end{figure}
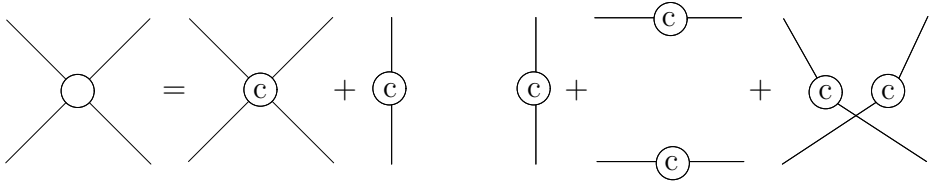

Consider the restricted class $\Gamma_c$ of Feynman graphs
that are connected. The $n$-point
functions $\Gconn^{(n)}$ defined by the corresponding restriction of
(\ref{eq:npfeyn}) are called the \emph{connected} $n$-point functions.
The relation between complete and connected $n$-point functions can be
described in a simple way. Partition the set of external legs of the
complete $n$-point function in all possible ways. The sum over the
product of connected functions for each partition yields the
complete $n$-point function,
\begin{equation}
 \Gcompl^{(n)}(x_1,\dots,x_n)=\sum_{k=1}^n
 \sum_{I_1\cup\cdots\cup I_k=\{x_1,\dots,x_n\}}
 \prod_{j=1}^k \Gconn(I_j) .
\label{eq:conncompl}
\end{equation}
Here $I_1,\dots,I_k$ denote non-empty subsets of $\{x_1,\dots,x_n\}$
forming a partition. Note also that the partitions are
\emph{unordered}, i.e., the subsets $I_1,\dots,I_k$ are not
distinguished. As an example, Figure~\ref{fig:compl4} shows the
decomposition of the complete $4$-point function in terms of
connected functions. The former is indicated by an empty circle, while
the latter are indicated by circles carrying the letter ``c''.

\begin{figure}
\setlength{\unitlength}{0.00083333in}
\begingroup\makeatletter\ifx\SetFigFont\undefined%
\gdef\SetFigFont#1#2#3#4#5{%
  \reset@font\fontsize{#1}{#2pt}%
  \fontfamily{#3}\fontseries{#4}\fontshape{#5}%
  \selectfont}%
\fi\endgroup%
{\renewcommand{\dashlinestretch}{30}
\begin{picture}(4595,265)(0,-10)
\path(12,121)(927,121)
\path(12,121)(927,121)
\path(2135,129)(1220,129)
\path(2135,129)(1220,129)
\path(3335,129)(2420,129)
\path(3335,129)(2420,129)
\path(4520,129)(3605,129)
\path(4520,129)(3605,129)
\put(447,71){\makebox(0,0)[lb]{{\SetFigFont{12}{14.4}{\rmdefault}{\mddefault}{\updefault}c}}}
\put(447,71){\makebox(0,0)[lb]{{\SetFigFont{12}{14.4}{\rmdefault}{\mddefault}{\updefault}c}}}
\put(484,113){\whiten\ellipse{210}{210}}
\put(484,113){\ellipse{210}{210}}
\texture{44555555 55aaaaaa aa555555 55aaaaaa aa555555 55aaaaaa aa555555 55aaaaaa 
	aa555555 55aaaaaa aa555555 55aaaaaa aa555555 55aaaaaa aa555555 55aaaaaa 
	aa555555 55aaaaaa aa555555 55aaaaaa aa555555 55aaaaaa aa555555 55aaaaaa 
	aa555555 55aaaaaa aa555555 55aaaaaa aa555555 55aaaaaa aa555555 55aaaaaa }
\put(2865,137){\shade\ellipse{212}{212}}
\put(2865,137){\ellipse{212}{212}}
\put(2865,137){\shade\ellipse{212}{212}}
\put(2865,137){\ellipse{212}{212}}
\put(4255,122){\shade\ellipse{212}{212}}
\put(4255,122){\ellipse{212}{212}}
\put(3875,122){\shade\ellipse{212}{212}}
\put(3875,122){\ellipse{212}{212}}
\put(995,69){\makebox(0,0)[lb]{{\SetFigFont{12}{14.4}{\rmdefault}{\mddefault}{\updefault}$=$}}}
\put(2195,69){\makebox(0,0)[lb]{{\SetFigFont{12}{14.4}{\rmdefault}{\mddefault}{\updefault}$+$}}}
\put(3380,69){\makebox(0,0)[lb]{{\SetFigFont{12}{14.4}{\rmdefault}{\mddefault}{\updefault}$+$}}}
\put(4595,69){\makebox(0,0)[lb]{{\SetFigFont{12}{14.4}{\rmdefault}{\mddefault}{\updefault}$+\,\,\cdots$}}}
\put(995,69){\makebox(0,0)[lb]{{\SetFigFont{12}{14.4}{\rmdefault}{\mddefault}{\updefault}$=$}}}
\put(2195,69){\makebox(0,0)[lb]{{\SetFigFont{12}{14.4}{\rmdefault}{\mddefault}{\updefault}$+$}}}
\put(3380,69){\makebox(0,0)[lb]{{\SetFigFont{12}{14.4}{\rmdefault}{\mddefault}{\updefault}$+$}}}
\put(4595,69){\makebox(0,0)[lb]{{\SetFigFont{12}{14.4}{\rmdefault}{\mddefault}{\updefault}$+\,\,\cdots$}}}
\put(447,61){\makebox(0,0)[lb]{{\SetFigFont{12}{14.4}{\rmdefault}{\mddefault}{\updefault}c}}}
\end{picture}
}
\caption{Decomposition of the connected propagator in terms of 1PI
  functions. Only part of the infinite sum is shown.}
\label{fig:conn2}
\end{figure}
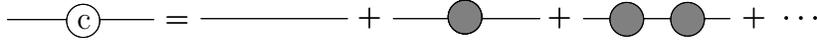

We turn to consider the restriction of the class of connected Feynman
graphs to that of \emph{1-particle irreducible (1PI)} Feynman
graphs, denoted by $\Gamma_{\text{1PI}}$. These are Feynman graphs
which are connected and remain connected when any one of their
internal edges is
cut. We define $\Gopi$ in analogy to (\ref{eq:npfeyn}) for
this restricted class. The relation between 1PI-functions and
connected ones may now be described as follows. The connected
$n$-point function is the sum over all tree graphs with the given
external legs, where the value of each vertex is given by
$\Gopi$. Since the latter carry Feynman propagators on all their legs,
internal edges connecting two vertices need to carry the inverse
Feynman propagator $\Gfeyn^{-1}$ to cancel one superfluous
Feynman propagator. As an example, Figure~\ref{fig:conn2} shows the
infinite decomposition of the connected $2$-point function
(propagator) in terms of 1PI ones. The 1PI vertices are drawn as
shaded discs.

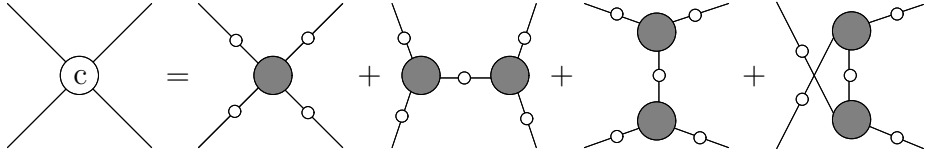
\begin{figure}
\setlength{\unitlength}{0.00083333in}
\begingroup\makeatletter\ifx\SetFigFont\undefined%
\gdef\SetFigFont#1#2#3#4#5{%
  \reset@font\fontsize{#1}{#2pt}%
  \fontfamily{#3}\fontseries{#4}\fontshape{#5}%
  \selectfont}%
\fi\endgroup%
{\renewcommand{\dashlinestretch}{30}
\begin{picture}(5739,954)(0,-10)
\texture{44555555 55aaaaaa aa555555 55aaaaaa aa555555 55aaaaaa aa555555 55aaaaaa 
	aa555555 55aaaaaa aa555555 55aaaaaa aa555555 55aaaaaa aa555555 55aaaaaa 
	aa555555 55aaaaaa aa555555 55aaaaaa aa555555 55aaaaaa aa555555 55aaaaaa 
	aa555555 55aaaaaa aa555555 55aaaaaa aa555555 55aaaaaa aa555555 55aaaaaa }
\path(2104,919)(1204,19)
\path(2104,919)(1204,19)
\path(1204,919)(2104,19)
\path(1204,919)(2104,19)
\path(3312,919)(3162,469)
\path(3312,919)(3162,469)
\path(2412,919)(2577,469)
\path(2412,919)(2577,469)
\path(2412,19)(2562,469)
\path(2412,19)(2562,469)
\path(3312,19)(3162,469)
\path(3312,19)(3162,469)
\path(3612,919)(4077,754)
\path(3612,919)(4077,754)
\path(4512,919)(4062,769)
\path(4512,919)(4062,769)
\path(3612,19)(4077,184)
\path(3612,19)(4077,184)
\path(4512,19)(4092,169)
\path(4512,19)(4092,169)
\path(5727,912)(5277,762)
\path(5727,912)(5277,762)
\path(5292,192)(5727,12)
\path(5292,192)(5727,12)
\path(4812,927)(5232,102)
\path(4812,927)(5232,102)
\path(4812,12)(5202,777)
\path(4812,12)(5202,777)
\path(12,919)(912,19)
\path(12,919)(912,19)
\path(912,919)(12,19)
\path(912,919)(12,19)
\path(2547,454)(3117,454)
\path(2547,454)(3117,454)
\path(4077,139)(4077,709)
\path(4077,139)(4077,709)
\path(5262,792)(5262,222)
\path(5262,792)(5262,222)
\put(1667,474){\ellipse{224}{224}}
\put(1669,469){\shade\ellipse{240}{240}}
\put(1669,469){\ellipse{240}{240}}
\put(2592,469){\shade\ellipse{240}{240}}
\put(2592,469){\ellipse{240}{240}}
\put(3147,469){\shade\ellipse{240}{240}}
\put(3147,469){\ellipse{240}{240}}
\put(4062,184){\shade\ellipse{240}{240}}
\put(4062,184){\ellipse{240}{240}}
\put(4062,739){\shade\ellipse{240}{240}}
\put(4062,739){\ellipse{240}{240}}
\put(5277,747){\shade\ellipse{240}{240}}
\put(5277,747){\ellipse{240}{240}}
\put(5277,192){\shade\ellipse{240}{240}}
\put(5277,192){\ellipse{240}{240}}
\put(461,469){\whiten\ellipse{238}{238}}
\put(461,469){\ellipse{238}{238}}
\put(1437,687){\whiten\ellipse{76}{76}}
\put(1437,687){\ellipse{76}{76}}
\put(1887,702){\whiten\ellipse{76}{76}}
\put(1887,702){\ellipse{76}{76}}
\put(1887,237){\whiten\ellipse{76}{76}}
\put(1887,237){\ellipse{76}{76}}
\put(1429,245){\whiten\ellipse{76}{76}}
\put(1429,245){\ellipse{76}{76}}
\put(3237,702){\whiten\ellipse{76}{76}}
\put(3237,702){\ellipse{76}{76}}
\put(2487,702){\whiten\ellipse{76}{76}}
\put(2487,702){\ellipse{76}{76}}
\put(2862,454){\whiten\ellipse{76}{76}}
\put(2862,454){\ellipse{76}{76}}
\put(3252,199){\whiten\ellipse{76}{76}}
\put(3252,199){\ellipse{76}{76}}
\put(2472,214){\whiten\ellipse{76}{76}}
\put(2472,214){\ellipse{76}{76}}
\put(4077,469){\whiten\ellipse{76}{76}}
\put(4077,469){\ellipse{76}{76}}
\put(4302,844){\whiten\ellipse{76}{76}}
\put(4302,844){\ellipse{76}{76}}
\put(3814,852){\whiten\ellipse{76}{76}}
\put(3814,852){\ellipse{76}{76}}
\put(4332,79){\whiten\ellipse{76}{76}}
\put(4332,79){\ellipse{76}{76}}
\put(3814,87){\whiten\ellipse{76}{76}}
\put(3814,87){\ellipse{76}{76}}
\put(5562,80){\whiten\ellipse{76}{76}}
\put(5562,80){\ellipse{76}{76}}
\put(5562,852){\whiten\ellipse{76}{76}}
\put(5562,852){\ellipse{76}{76}}
\put(4969,620){\whiten\ellipse{76}{76}}
\put(4969,620){\ellipse{76}{76}}
\put(4969,320){\whiten\ellipse{76}{76}}
\put(4969,320){\ellipse{76}{76}}
\put(5269,469){\whiten\ellipse{76}{76}}
\put(5269,469){\ellipse{76}{76}}
\put(1077,409){\makebox(0,0)[b]{{\SetFigFont{12}{14.4}{\rmdefault}{\mddefault}{\updefault}$=$}}}
\put(2269,402){\makebox(0,0)[b]{{\SetFigFont{12}{14.4}{\rmdefault}{\mddefault}{\updefault}$+$}}}
\put(3469,402){\makebox(0,0)[b]{{\SetFigFont{12}{14.4}{\rmdefault}{\mddefault}{\updefault}$+$}}}
\put(424,417){\makebox(0,0)[lb]{{\SetFigFont{12}{14.4}{\rmdefault}{\mddefault}{\updefault}c}}}
\put(4669,402){\makebox(0,0)[b]{{\SetFigFont{12}{14.4}{\rmdefault}{\mddefault}{\updefault}$+$}}}
\end{picture}
}
\caption{Decomposition of the connected $4$-point function in terms of
modified 1PI functions.}
\label{fig:conn4}
\end{figure}

One variant of the 1PI-functions is of interest. Namely, replace
the Feynman propagator $\Gfeyn$ on the external legs of a
1PI-function by the connected propagator $\Gconn^{(2)}$. We denote these
modified 1PI-functions by $\Gopim$. The relation between $\Gconn$
and $\Gopim$ is the same as that between $\Gconn$ and
$\Gopi$ described above, except that internal edges now carry
the inverse $(\Gconn^{(2)})^{-1}$ of the connected propagator (defined
analogous to
(\ref{eq:invfey})). A crucial property of the modified 1PI-functions
is that the $2$-point function $\Gopim^{(2)}$ vanishes by
definition. This
means that only trees with vertices that have valence at least three
can occur in the sum. In particular, this makes the sum over trees
finite for any given set of external legs. Figure~\ref{fig:conn4}
shows the decomposition of the connected $4$-point function in terms
of modified 1PI functions. The fact that the legs now carry the
connected propagator is indicated with little circles.

Note that we assume all 1-point functions to vanish.

\subsection{Field operator algebra and coproduct}

We turn to introduce the basic algebraic definitions and elementary
formalism employed in this article. While our basic setup is
largely the same
as that in \cite{BFFO:twist} we give an adapted and
self-contained description here.

Let $V$ be the vector space of linear combinations of elementary field
operators $\phi(x)$. That is, elements of
$V$ take the form $\lambda_1 \phi(x_1)+\lambda_2 \phi(x_2)+\cdots
+\lambda_n \phi(x_n)$, where $\lambda_i$ are complex
numbers and $x_i$ denote field operator labels.
Consider now the commutative algebra $\SV$ generated by those field
operators with the \emph{time-ordered} product.\footnote{In
  \cite{BFFO:twist} the same algebra was considered, but with the
  \emph{normal ordered} product. This makes no difference to its
  structure. It is just more convenient in the present context to
  start immediately with time-ordered product.} A general element in
$\SV$ takes the form
\[
 \lambda_1 \phi(x_{1,1})\phi(x_{1,2})\cdots\phi(x_{1,k_1})
 +\lambda_2 \phi(x_{2,1})\phi(x_{2,2})\cdots\phi(x_{2,k_2})
 +\cdots .
\]
Note that we do not explicitly indicate the time-ordering prescription,
but it is always understood. Let us denote by $V^k$ the vector space of
$k$-fold products of field operator. Then, $\SV$ is the direct sum
of the spaces $V^k$, i.e.,
$\SV=\bigoplus_{k=0}^\infty V^k$.
We denote the identity operator spanning $V^0$ by $\one$.

In mathematical terms,
$\SV$ is called the \emph{symmetric algebra} over $V$.
In our current notation a time-ordered
$n$-point function is a linear map $V^n\to \C$. Thus, the ensemble of
time-ordered
$n$-point functions (of a given type) determines a linear map
$\SV\to \C$. Recalling the various types of $n$-point functions
introduced in the previous section we denote the corresponding linear
maps $\SV\to\C$ as follows,
\begin{align*}
\compl(\phi(x_1)\cdots\phi(x_n)) & \defeq \Gcompl^{(n)}(x_1,\dots,x_n)\\
\conn(\phi(x_1)\cdots\phi(x_n)) & \defeq \Gconn^{(n)}(x_1,\dots,x_n)\\
\opi(\phi(x_1)\cdots\phi(x_n)) & \defeq \Gopi^{(n)}(x_1,\dots,x_n)\\
\opim(\phi(x_1)\cdots\phi(x_n)) & \defeq \Gopim^{(n)}(x_1,\dots,x_n) .
\end{align*}
The assumption that all 1-point functions vanish means that
$\compl(\phi(x))=\conn(\phi(x))=\opi(\phi(x))=\opim(\phi(x))=0$.
For completeness, we also need to define 0-point functions. Since
$\compl(\one)=\langle 0| 0\rangle$ we need to set $\compl(\one)=1$. We
shall see that the consistent choice for the other 0-point functions is
$\conn(\one)=\opi(\one)=\opim(\one)=0$. Also, the 2-point function
$\opim(\phi(x)\phi(y))$ vanishes by construction.

$\SV$ is not only an algebra, but also a \emph{coalgebra} in a
natural way. This means that there exists a linear \emph{coproduct} map
$\cop:\SV\to \SV\tens \SV$ with certain properties. One
way to think about this coproduct map is as a way to split a product of
field operators into two parts in all possible ways. For example,
\begin{align}
\cop(\one) = & \one\tens\one, \label{eq:copzero} \\
\cop(\phi(x)) = & \phi(x)\tens \one + \one\tens\phi(x),
\label{eq:copone}\\
\begin{split}
\cop(\phi(x)\phi(y)) = &\phi(x)\phi(y)\tens\one
 + \phi(x)\tens\phi(y) \\
 & + \phi(y)\tens\phi(x) + \one\tens\phi(x)\phi(y) .
\label{eq:coptwo}
\end{split}
\end{align}
The general formula for the coproduct is
\begin{equation}
\cop (\phi(x_1)\cdots \phi(x_n))=
 \sum_{I_1\cup I_2 = \{\phi(x_1),\dots,\phi(x_n)\}} T(I_1)\tens T(I_2) .
\label{eq:coproduct}
\end{equation}
Here the sum runs over partitions of the set of field operators
$\{\phi(x_1),\dots,\phi(x_n)\}$ into two sets $I_1$ and $I_2$. $T$
denotes the time-ordered product of the field operators in the
corresponding partition. 

The coproduct has the property that it is an algebra map. This means
that $\cop(\phi(x_1)\cdots\phi(x_n))=\cop(\phi(x_1)\cdots\phi(x_k))
\cdot\cop(\phi(x_{k+1})\cdots\phi(x_n))$. Here, the product in
$\SV$ is extended to a product in $\SV\tens\SV$ in the
obvious way. The property of the coproduct to be an algebra
map together with (\ref{eq:copzero}) and (\ref{eq:copone}) completely
determines it. Formula (\ref{eq:coproduct}) can be derived from these
properties.

Another important structure is the \emph{counit}
$\cou:\SV\to\C$. It is defined by $\cou(\one)=1$ and
$\cou(\phi(x_1)\cdots\phi(x_n))=0$ for $n>0$.
The characterizing property of the counit is the equality
$(\cou\tens\id)\circ\cop=\id=(\id\tens\cou)\circ\cop$.
The algebra $\SV$ together with unit, counit, coproduct and
antipode (which is another map that we do not need here) forms a
\emph{Hopf algebra}, see \cite{BFFO:twist}.

We shall also need iterated coproducts. First note that the coproduct
satisfies the equality
$(\cop\tens\id)\circ\cop=(\id\tens\cop)\circ\cop$. That is, after
applying the coproduct once, a second application either on the first
or on the second component yield the same result.
This is called
\emph{coassociativity}. We define the map $\cop^k:\SV\to
\SV^{\tens k+1}$ as the $k$-fold application of the coproduct. Here,
$\SV^{\tens k+1}$ denotes the $(k+1)$-fold tensor product of
$\SV$. Thus, $\cop^0=\id$ and
$\cop^{k+1}=(\cop\tens\id^{\tens k})\circ\cop^k$. Here, the latter
equation
could be written in $k+1$ different ways (corresponding to different
positions of the application of the coproduct) which are all
equivalent due to coassociativity. The map $\cop^k$ generalizes
(\ref{eq:coproduct}) as follows:
\begin{equation}
\cop^k (\phi(x_1)\cdots \phi(x_n))=
 \sum_{I_1\cup \cdots\cup I_{k+1} = \{\phi(x_1),\dots,\phi(x_n)\}}
 T(I_1)\tens \cdots \tens T(I_{k+1}) .
\label{eq:copk}
\end{equation}
The difference to the single coproduct is that the set of field
operators is now split into $k+1$ partitions. Note also that the
partitions are \emph{ordered}, i.e., the sets $I_1\dots I_{k+1}$ are
distinguishable.

A further definition we will require is the following. Given linear maps
$\alpha:\SV\to\C$ and $\beta:\SV\to\C$ their \emph{convolution
  product} is the map $\alpha\star\beta:\SV\to\C$ defined by
$(\alpha\tens\beta)\circ\cop$. That is, we apply the coproduct
followed by the application of $\alpha$ to its first and $\beta$ to
its second component. Note that the coassociativity of the coproduct
implies associativity of the convolution product. Thus, we can write
multiple convolution products without needing to specify brackets. In
particular, we may write an iterated convolution product using the
iterated coproduct,
\begin{equation}
\alpha_1\star\cdots\star\alpha_k
=(\alpha_1\tens\cdots\tens\alpha_{k})\circ\cop^{k-1} .
\label{eq:convprodk}
\end{equation}
The convolution product makes $(\SV)^*$, the space of complex
linear functions on $\SV$, into an algebra. This
algebra has a unit which is given by the counit $\cou$ of
$\SV$. Furthermore, any element $\alpha$ which satisfies
$\alpha(\one)\neq 0$ is invertible in this algebra, i.e., has an
inverse with respect to the convolution product.

\section{Complete and connected $n$-point functions}
\label{sec:complconn}

The first instance where we shall apply the Hopf algebraic
approach to capture the combinatorics of quantum field theory is in
the relation between the complete and the connected $n$-point
functions.

As the (iterated) coproduct (\ref{eq:copk}) is intimately related to
partitions it seems predestined to express the relation between
complete and connected $n$-point functions
(\ref{eq:conncompl}). Indeed, the relation between the two is very
compactly and elegantly expressed using the convolution product (and
thus implicitly the coproduct).

\begin{prop}
\label{prop:exp}
The complete $n$-point functions may be expressed in terms of the
connected ones through the convolution exponential, $\compl=\exp_\star
\conn$. The convolution exponential is defined in terms of its
power series expansion.
\end{prop}
\begin{proof}
We explicitly perform the power series expansion on a given argument
in the subspace $V^n\subset\SV$ ($n>0$),
\begin{equation*}
\begin{split}
(\exp_\star(\conn))(\phi(x_1)\cdots\phi(x_n))
& = \sum_{k=0}^{\infty}\,\frac{1}{k!}\conn^{\star k}
 (\phi(x_1)\cdots\phi(x_n))\\
& =
\sum_{k=1}^{\infty}\,\frac{1}{k!}(\conn\tens\cdots\tens\conn)
\circ\cop^{k-1} (\phi(x_1)\cdots\phi(x_n))\\
& = \sum^{n}_{k=1}\,\frac{1}{k!}
\sum_{I_1\cup \cdots\cup I_{k} = \{\phi(x_1),\dots,\phi(x_n)\}}
\prod_{j=1}^k \conn(T(I_j)) .
\end{split}
\end{equation*}
Here, $\conn^{\star k}$ denotes the $k$-fold convolution product of
$\conn$ with itself. In particular, $\conn^{\star 0}=\cou$ by
definition. Since $\cou(\phi(x_1)\cdots\phi(x_n))=0$ the first summand
is zero in the first line and we may omit it. In going from the second
to the third line we insert the
definition of the iterated coproduct (\ref{eq:copk}). Note that by
definition $\conn(\one)=0$. This implies that all partitions where at
least one partitioning set $I_j$ is empty do not contribute. In
particular, all summands of the outer sum with $k>n$ must vanish. The
only difference to the partitioning in (\ref{eq:conncompl}) is that
the latter are unordered. However, the number of occurrences of each
unordered partition in the set of ordered ones is exactly $k!$. Thus,
the factor $1/k!$ establishes equality with
$\compl(\phi(x_1)\cdots\phi(x_n))$. To complete the proof, note
that $\compl(\one)=\conn^{\star 0}(\one)=\cou(\one)=1$ since
$\conn(\one)=0$ and thus $\conn^{\star k}(\one)=0$ for $k>0$.
\end{proof}

Let us emphasize that although the power series defining the exponential
is formally infinite, on any given element of $\SV$ it is
truncated to a finite and thus well defined sum as shown in the proof
above. Thus, the relation $\compl=\exp_\star \conn$ is completely well
defined algebraically. Indeed, we may even invert it.

\begin{cor}
The connected $n$-point functions may be expressed in terms of the
complete ones through the convolution logarithm, $\conn=\log_\star
\compl$. The convolution logarithm is defined in terms of its power
series expansion in $\compl-\cou$.
\end{cor}
\begin{proof}
By definition
\[
\log_\star\compl=\sum_{k=0}^\infty (-1)^k\frac{1}{k}
  (\compl-\cou)^{\star k} .
\]
Note that $(\compl-\cou)(\one)=0$. This implies as in the proof of
Proposition~\ref{prop:exp} that the sum truncates to a finite sum on any
given element in $\SV$. One may show that the operations
$\exp_\star$ and $\log_\star$ are mutually inverse by inserting one
power series into the other. Since the procedure and result is exactly
the same as in the usual arithmetic of complex numbers (say) we do not
perform it explicitly here. However, it is crucial that the truncation of
the power series for any given argument in $\SV$ makes it
algebraically well defined in the present context.
\end{proof}

The attentive reader will notice that the present method of relating the
complete and connected $n$-point functions shows certain similarities to
the conventional one. Namely, in the conventional method one also finds
that the relation is given by the exponential and the logarithm
respectively, see e.g.\ \cite{ItZu:qft}. The difference is of course
that the conventional method uses sources while the present one uses
the coproduct to define exponential and logarithm.

\section{Connected and 1PI $n$-point functions}
\label{sec:conn1pi}

We now turn to the relation between connected and 1PI $n$-point
functions. To state our Hopf algebraic formulation of this relation we
need to define a few auxiliary structures first.

Define the formal
element $\Rop\in \SV\tens\SV$ using the inverse
Feynman propagator (\ref{eq:invfey}) as follows,\footnote{$R$ is
  formal insofar as it really lives in
a completion of the tensor product $\SV\tens\SV$. However,
this fact is largely irrelevant for our purposes.}
\begin{equation}
\Rop \defeq \int
\xd x\, \xd y\,G_F^{-1}(x,y)\,(\phi(x)\tens\phi(y)) .
\label{eq:R}
\end{equation}
Using the product in $\SV\tens\SV$ we may view $\Rop$ as an
operator acting on this space by multiplication. Define now the map
$\Qop:\SV\to \SV\tens\SV$ as the composition of $\Rop$
with the
coproduct together with a factor of $1/2$:
\begin{equation}
\Qop \defeq \frac{1}{2} \Rop\circ\cop .
\label{eq:Q}
\end{equation}
We generalize $\Qop$ to maps $\Qop_i:\SV^{\tens k}\to
 \SV^{\tens k+1}$.
Namely, let $\Qop_i$ be the application of $Q$ on the
$i^{\mbox{\tiny{th}}}$ component only, i.e.,
\begin{equation*}
\Qop_i\defeq\id^{\tens i-1}\tens \Qop\tens \id^{\tens k-i} .
\end{equation*}
Finally, we define maps $\Dop^k:\SV\to \SV^{\tens k+1}$
for $k\in\N_0$ recursively as follows,
\begin{align}
 \Dop^0 & \defeq \id, \nonumber \\
 \Dop^k & \defeq \frac{1}{k} \sum_{i=1}^k Q_i \circ \Dop^{k-1} .
 \label{eq:reclambda}
\end{align}

We are now ready to state our main result.
\begin{thm}
\label{thm:conn1pi}
The connected $n$-point functions may be expressed in terms of the 1PI
ones through the formula
\begin{equation}
\conn=\sum_{k=1}^\infty \conn^k,\quad\text{with}\quad
 \conn^k\defeq \opi^{\tens k}\circ\Dop^{k-1} .
\label{eq:conn1pi}
\end{equation}
\end{thm}
The remainder of this section will be devoted to the proof of this
result.

Recall from the description in Section~\ref{sec:review} that the
connected $n$-point functions are expressible as the sum over all tree
graphs with 1PI $n$-point functions as vertices. Since the latter
are represented by $\opi$ we see that the sum in formula
(\ref{eq:conn1pi}) must
correspond to the sum over the number of such vertices $k$. In turn,
the map $\Dop^k$ must contain the information about all tree
graphs with $k$ vertices. We proceed to explain this.

We first generalize the definition of $\Rop$ given in
(\ref{eq:R}) to an element (or operator) $\Rop_{i,j}\in\SV^{\tens
  k}$ with $1\le i<j\le k$ via
\begin{equation}\label{eq:Rij}
\Rop_{i,j} := \int
\xd x\,\xd y\,G_F^{-1}(x,y)\,(\one^{\tens i-1}\tens \phi(x)
\tens\one^{\tens j-i-1}
\tens\phi(y)\tens\one^{\tens k-j}) .
\end{equation}
In other words, the field operators $\phi(x)$ and $\phi(y)$ are
inserted at the $i^{\mbox{\tiny{th}}}$ and
$j^{\mbox{\tiny{th}}}$ position respectively.

\begin{figure}
\setlength{\unitlength}{0.00083333in}
\begingroup\makeatletter\ifx\SetFigFont\undefined%
\gdef\SetFigFont#1#2#3#4#5{%
  \reset@font\fontsize{#1}{#2pt}%
  \fontfamily{#3}\fontseries{#4}\fontshape{#5}%
  \selectfont}%
\fi\endgroup%
{\renewcommand{\dashlinestretch}{30}
\begin{picture}(5412,2239)(0,-10)
\path(1606,1963)(1846,1438)
\path(1516,1018)(1831,1543)
\path(2386,1408)(1771,1498)
\put(2491,1318){\makebox(0,0)[lb]{{\SetFigFont{12}{14.4}{\rmdefault}{\mddefault}{\updefault}$x_2$}}}
\put(1261,823){\makebox(0,0)[lb]{{\SetFigFont{12}{14.4}{\rmdefault}{\mddefault}{\updefault}$x_3$}}}
\put(1366,2068){\makebox(0,0)[lb]{{\SetFigFont{12}{14.4}{\rmdefault}{\mddefault}{\updefault}$x_1$}}}
\texture{44555555 55aaaaaa aa555555 55aaaaaa aa555555 55aaaaaa aa555555 55aaaaaa 
	aa555555 55aaaaaa aa555555 55aaaaaa aa555555 55aaaaaa aa555555 55aaaaaa 
	aa555555 55aaaaaa aa555555 55aaaaaa aa555555 55aaaaaa aa555555 55aaaaaa 
	aa555555 55aaaaaa aa555555 55aaaaaa aa555555 55aaaaaa aa555555 55aaaaaa }
\path(4974,1940)(4824,1490)
\path(4974,1940)(4824,1490)
\path(4074,1940)(4239,1490)
\path(4074,1940)(4239,1490)
\path(4074,1040)(4224,1490)
\path(4074,1040)(4224,1490)
\path(4974,1040)(4824,1490)
\path(4974,1040)(4824,1490)
\path(4209,1475)(4779,1475)
\path(4209,1475)(4779,1475)
\put(1815,1490){\shade\ellipse{240}{240}}
\put(1815,1490){\ellipse{240}{240}}
\put(4254,1490){\shade\ellipse{240}{240}}
\put(4254,1490){\ellipse{240}{240}}
\put(4809,1490){\shade\ellipse{240}{240}}
\put(4809,1490){\ellipse{240}{240}}
\put(128,1468){\shade\ellipse{240}{240}}
\put(128,1468){\ellipse{240}{240}}
\put(91,66){\makebox(0,0)[lb]{{\SetFigFont{12}{14.4}{\rmdefault}{\mddefault}{\updefault}$\one$}}}
\put(1261,66){\makebox(0,0)[lb]{{\SetFigFont{12}{14.4}{\rmdefault}{\mddefault}{\updefault}$\phi(x_1)\phi(x_2)\phi(x_3)$}}}
\put(3256,58){\makebox(0,0)[lb]{{\SetFigFont{12}{14.4}{\rmdefault}{\mddefault}{\updefault}$\Rop\cdot(\phi(x_1)\phi(x_2)\otimes\phi(x_3)\phi(x_4))$}}}
\put(4951,2023){\makebox(0,0)[lb]{{\SetFigFont{12}{14.4}{\rmdefault}{\mddefault}{\updefault}$x_3$}}}
\put(4944,838){\makebox(0,0)[lb]{{\SetFigFont{12}{14.4}{\rmdefault}{\mddefault}{\updefault}$x_4$}}}
\put(3946,2023){\makebox(0,0)[lb]{{\SetFigFont{12}{14.4}{\rmdefault}{\mddefault}{\updefault}$x_1$}}}
\put(3946,823){\makebox(0,0)[lb]{{\SetFigFont{12}{14.4}{\rmdefault}{\mddefault}{\updefault}$x_2$}}}
\end{picture}
}
\caption{Examples for the correspondence of graphs with elements of
 $\SV^{\tens k}$.}
\label{fig:exacor}
\end{figure}
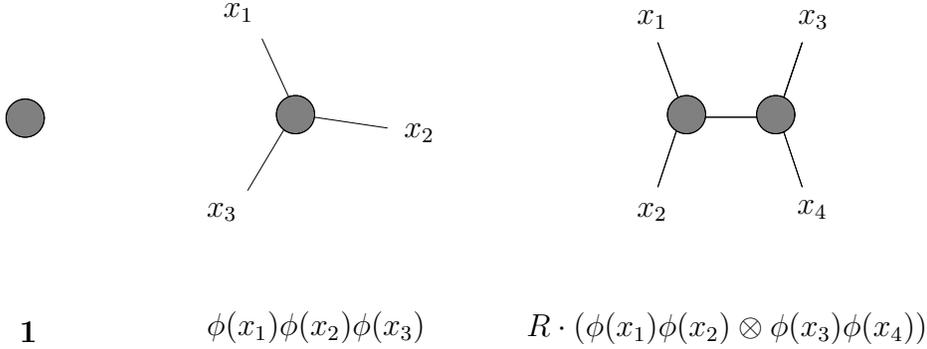

We proceed to establish a correspondence between graphs with $k$
vertices and certain elements of $\SV^{\tens k}$.
Each tensor factor of $\SV^{\tens k}$ corresponds to one
vertex. A product $\phi(x_1)\cdots\phi(x_n)$ in a given tensor factor
corresponds to external legs labeled by $x_1,\dots,x_n$. The element
$\Rop_{i,j}\in \SV^{\tens k}$ corresponds to an internal edge
connecting the $i^{\mbox{\tiny{th}}}$ vertex with the
$j^{\mbox{\tiny{th}}}$ vertex. Combining several internal edges and
external legs by multiplying the respective expressions in
$\SV^{\tens k}$ allows to build arbitrary graphs with $k$
vertices. Figure~\ref{fig:exacor} shows some examples. Applying
$\opi^{\tens k}$ to the resulting expression obviously yields the
value of the respective graph with the vertices being 1PI
functions. Thus, the graphs we just discussed are exactly those
that are to enter into expressing the connected functions in
terms of the 1PI ones.

The ordering of the tensor factors of $\SV^{\tens k}$
induces an ordering of the vertices of the graph, i.e., we may think
of them as labeled with numbers $1,\dots,k$.
However, when applying
$\opi^{\tens k}$ the ordering is ``forgotten''. Indeed, it is not
relevant for the interpretation of graphs, but only plays a role at
the level of their algebraic representation here. For short, we call a
graph \emph{ordered} if its vertices are ordered.
In the following, we will encounter elements of $\SV^{\tens k}$
that are linear combinations of expressions corresponding to ordered
graphs. Alternatively, we might think of such elements as linear
combinations of
unordered graphs by considering different ordered graphs that
correspond to the same unordered graph as the same.
We call \emph{weight} of the graph 
the scalar multiplying the expression for a given graph.
Clearly, the weight of an unordered graph is the sum of the weights of
all corresponding ordered graphs.

We see now what is required to prove
Theorem~\ref{thm:conn1pi}. Namely, we need to show that
$\Dop^{k-1}(\phi(x_1)\cdots\phi(x_n))\in\SV^{\tens k}$ corresponds
exactly to the sum over all tree graphs with $k$ (unordered!) vertices
and external legs labeled by $x_1,\dots,x_n$, each with weight one.
Actually, we need to show something slightly weaker, namely, the
statement needs to apply only to graphs where all of the vertices have
valence at least two. This is because the 0-point and 1-point
1PI functions are zero. Indeed, we will see in the process that the
more general statement is false.

Consider the coproduct applied to the
$i^{\mbox{\tiny{th}}}$ component of $\SV^{\tens k}$, i.e.,
$\cop_i\defeq \id^{\tens i-1}
\tens\cop\tens\id^{\tens k-i}$ as a map $\SV^{\tens k}\to
\SV^{\tens k+1}$.
Recalling the formula (\ref{eq:coproduct}) we see that
$\cop_i$ converts a graph with $k$ vertices into a sum over
graphs with $k+1$ vertices by \emph{splitting} the
$i^{\mbox{\tiny{th}}}$ vertex into two in all possible ways. That is,
the $i^{\mbox{\tiny{th}}}$ vertex is replaced by two vertices
(numbered $i$ and $i+1$) and its
legs (considered as distinguishable)
are distributed between the two new vertices in all possible ways.
Note that the two new vertices are distinguished due to the ordering
of the tensor factors. Thus, to obtain the corresponding
operation for unordered graphs we need to divide by a factor of $2$.
This factor corresponds to the two different relative orderings of the
new vertices with which each unordered configuration occurs.
The only exception to this is the case when the split vertex has no
legs at all. No overcounting happens in this case.
The meaning of the map $\Qop_i$ given by (\ref{eq:Q})
becomes clear now in terms of graphs. Namely, it splits the
$i^{\mbox{\tiny{th}}}$ vertex into two and subsequently reconnects the
two new vertices with an edge. Dividing by $2$ compensates for the
double counting
as described above if we are interested in unordered graphs (assuming
the set of legs of the split vertex is not empty).

\begin{lem}
\label{lem:alltrees}
Fix integers $k\ge 1$ and $n\ge 0$ as well as field operator labels
$x_1,\dots,x_n$.
(a) $\Dop^{k-1}(\phi(x_1)\cdots\phi(x_n))$
corresponds in the manner
described above to a sum of weighted graph with
$k$ vertices and $n$ external legs labeled by
$x_1,\dots,x_n$. (b) Each of these graphs is connected.
(c) Each of these graphs is a tree graph. (d) Any tree graph with $k$
unlabeled vertices and the given external legs occurs
in $\Dop^{k-1}(\phi(x_1)\cdots\phi(x_n))$ with some positive weight.
\end{lem}
\begin{proof}
Firstly, it is clear that $\Dop^0(\phi(x_1)\cdots\phi(x_n))$
corresponds to the graph with one vertex and the external legs
labeled by $x_1,\dots,x_n$. Secondly,
$\Dop^{k-1}(\phi(x_1)\cdots\phi(x_n))$ is generated from this by sums
of multiple applications of maps $\Qop_i$ and scalar factors. But
$\Qop_i$ converts a
term corresponding to a graph to a sum over terms corresponding to
graphs. Thus,
$\Dop^{k-1}(\phi(x_1)\cdots\phi(x_n))$ is a sum of terms each of which
corresponds to a graph (with some weight). This completes the proof of
(a).

Given a connected graph, splitting a vertex produces at most two
disconnected pieces. Reconnecting the new vertices with an edge yields
again a
connected graph. Thus, $\Qop_i$ produces connected graphs from
connected ones. Evidently, $\Dop^0(\phi(x_1)\cdots\phi(x_n))$
corresponds to a connected graph. This
proves (b).

Splitting a vertex of a tree graph necessarily yields two
disconnected graphs. Reconnecting the new vertices with an edge thus
cannot
introduce a cycle. Therefore, $\Qop_i$ produces tree graphs from tree
graphs. Evidently, $\Dop^0(\phi(x_1)\cdots\phi(x_n))$ corresponds to
a tree graph. This proves (c).

To prove (d) we use a recursive argument. Evidently, for $k=1$ the
statement is true. Now assume that any tree graph with $k-1$ unlabeled
vertices and the given external legs occurs in
$\Dop^{k-2}(\phi(x_1)\cdots\phi(x_n))$ with positive weight.
Consider
a tree graph with $k$ unlabeled vertices and the given external
legs. Choose an arbitrary internal edge. Shrinking this edge and
fusing the vertices it connects yields a tree graph that corresponds
by assumption to a term in
$\Dop^{k-2}(\phi(x_1)\cdots\phi(x_n))$. Say, the
fused vertex has position $i$. Applying $\Qop_i$ to this term will
yield a sum over terms one of which will correspond to the original
tree graph. By the recursive definition of
$\Dop^{k-1}(\phi(x_1)\cdots\phi(x_n))$ it thus contains this term with
positive weight. This completes the proof.
\end{proof}

What remains in order to prove Theorem~\ref{thm:conn1pi} is to show
that the term corresponding to each tree graph (with all vertices of
valence at least two) has weight exactly $1$. We start with a more
restricted result.
\begin{lem}
\label{lem:weightwlegs}
Fix integers $k\ge 1$ and $n\ge k$ and field operator
labels $x_1,\dots,x_n$. Consider a tree with $k$ vertices, external
legs labeled by $x_1,\dots,x_n$ and the property that each vertex has
at least one external leg. Then, the term in
$\Dop^{k-1}(\phi(x_1)\cdots\phi(x_n))$ corresponding to that tree has
weight $1$.
\end{lem}
\begin{proof}
We proceed by induction on the number of vertices. Clearly, the
statement is true for $k=1$. Now assume that
$\Dop^{k-2}(\phi(x_1)\cdots\phi(x_n))$ contains each tree graph with
$k-1$ vertices and external legs labeled by $x_1,\dots,x_n$ and the
property that each vertex carries at least one external leg with
weight exactly $1$. (Of course it may in addition contain terms
corresponding to other graphs.)
Consider a tree graph $\gamma$ with $k$ vertices, external legs labeled by
$x_1,\dots,x_n$ and the property that each vertex carries at least one
external leg. We proceed to show that it occurs with weight exactly
$1$ in $\Dop^{k-1}(\phi(x_1)\cdots\phi(x_n))$.
To this end we check from which graphs with $k-1$ vertices $\gamma$ is
generated by the recursion formula (\ref{eq:reclambda}) and in how
many ways. Translated into the language of graphs, the
formula prescribes that a given graph is split and reconnected at
every of its vertices. The resulting terms are summed over and
multiplied by $1/k$. Conversely, this implies that $\gamma$ is
generated from all the graphs that are obtained by shrinking one of
its internal edges. Since there are $k$ edges these are a priori $k$
graphs. These graphs are indeed all distinct, since the external legs
attached to each vertex force them to be distinguishable. Furthermore,
each of these graphs generates $\gamma$ in only one way, i.e., as one
resulting graph in the
splitting and reconnecting of only one of its vertices (again due to
the forced distinguishability of the vertices). Since by assumption
each generating graph has weight one, the multiplicity $k$ cancels
exactly with the factor $1/k$ in (\ref{eq:reclambda}) to produce
weight $1$ for $\gamma$. This completes the proof.
\end{proof}

To describe the weight of arbitrary tree graphs we will need to
consider symmetries of graphs.
\begin{dfn}
\label{def:symfac}
Consider a tree graph $\gamma$ with ordered vertices. A
\emph{symmetry} of $\gamma$ is a permutation of the ordering of its
vertices that
yields the (topologically) same ordered graph. The number of
symmetries, i.e., order
of the group of permutations leaving the graph invariant, is called the
\emph{symmetry factor} of the graph.
\end{dfn}
Since the symmetry factor is the same for any ordering of the vertices
of a graph the concept makes sense for unordered graphs as well.

We also need the following property of $\Dop^k$:
\begin{lem}
\label{lem:lambdafac}
Fix integers $k,n\ge 0$ and operator labels $x_1,\dots,x_n$. Then,
$\Dop^k$ satisfies the factorization property
\begin{equation*}
\Dop^k(\phi(x_1)\cdots\phi(x_n))=\Dop^k(\one)\cdot
 \cop^k(\phi(x_1)\cdots\phi(x_n)).
\end{equation*}
\end{lem}
\begin{proof}
This follows immediately from the multiplicativity of the coproduct
and the recursive definition (\ref{eq:reclambda}).
\end{proof}

We can now state the generalization of Lemma~\ref{lem:weightwlegs}.
\begin{lem}
\label{lem:weightgeneral}
Fix integers $k\ge 1$ and $n\ge 0$ and field operator
labels $x_1,\dots,x_n$. Consider a tree $\gamma$ with $k$ vertices and
external legs labeled by $x_1,\dots,x_n$. Let $s$ be the symmetry
factor of $\gamma$. Then, the term in 
$\Dop^{k-1}(\phi(x_1)\cdots\phi(x_n))$ corresponding to that tree has
weight $1/s$.
\end{lem}
\begin{proof}
If $\gamma$ has external legs attached to every of its vertices we
simply recall Lemma~\ref{lem:weightwlegs} and note that a graph with
this property has no non-trivial symmetries. Thus, we may now assume
that $\gamma$ has $m$ vertices to which no external leg is
attached.
Consider a graph
$\gamma'$ which is constructed from $\gamma$ by attaching an external
leg to every vertex without external legs, choosing arbitrary but
fixed labels $y_1,\dots,y_m$ for the legs in the process.
By
Lemma~\ref{lem:lambdafac} and the multiplicativity of $\cop$ and thus
$\cop^j$ we have
\begin{multline*}
\Dop^{k-1}(\phi(x_1)\cdots\phi(x_n)\phi(y_1)\cdots\phi(y_m))=\\
 \Dop^{k-1}(\phi(x_1)\cdots\phi(x_n))\cdot
 \cop^{k-1}(\phi(y_1)\cdots\phi(y_m)) .
\end{multline*}
By Lemma~\ref{lem:weightwlegs}, the graph $\gamma'$ occurs in the
term on left hand side with weight $1$. By Lemma~\ref{lem:alltrees},
the graph $\gamma$ occurs in the
first factor on the right hand side with some non-zero weight, say
$\alpha$.
Every summand of $\cop^{k-1}(\phi(y_1)\cdots\phi(y_m))$ (recall the
formula (\ref{eq:copk}))
which places the external legs at the designated vertices of $\gamma$
to produce $\gamma'$ contributes to the weight of $\gamma'$ in terms
of that of $\gamma$. Any different ways this can happen define a
symmetry of $\gamma$. Furthermore, $\gamma$ can have no more than
these symmetries, since its vertices that already carry external legs
are distinguishable and thus held fixed under any symmetry.
Therefore, we obtain the formula $1=\alpha\cdot s$
for the weights, or $\alpha=1/s$. This completes the proof.
\end{proof}

The appendix shows the result of computing all tree graphs without
external legs as weighted contributions to
$\Dop^{k-1}(\one)$, for vertex number $k\le 7$.

The proof of Theorem~\ref{thm:conn1pi} is completed with the following
lemma.
\begin{lem}
Consider a tree graph $\gamma$, all of whose vertices have valence at
least two. Then, $\gamma$ has no non-trivial symmetries.
\end{lem}
\begin{proof}
Consider a vertex $v$ of $\gamma$. We show that any symmetry must
leave $v$ invariant. If $v$ carries an external leg it must be
invariant since it is distinguishable. Thus, assume $v$ carries no
external leg. Choose one internal edge $e$ connected to $v$. Cut
$\gamma$ into
two by removing $e$. This yields two tree graphs $\gamma_1$ and
$\gamma_2$. Each of these must have at least one external leg to
satisfy the
valence requirement. Say $e_1$ is an external leg of $\gamma_1$ and
$e_2$ an external leg of $\gamma_2$. Since $\gamma$ is a tree there is
exactly one path to connect $e_1$ with $e_2$. Since the vertices
connected with $e_1$ and $e_2$ are held fixed under any symmetry so is
the whole chain of vertices formed by the path. However, $v$ is part
of this chain by construction and thus held fixed by any
symmetry.
\end{proof}

\section{Further recursion relations}
\label{sec:recrel}

We may extend the results of the previous section to obtain further
interesting recursion relations. Recall from (\ref{eq:conn1pi})
the decomposition of $\conn$
into components $\conn^k$ according to vertex number $k$.
\begin{prop}
\label{prop:recconn}
$\conn^k$ may be determined recursively via $\conn^1=\opi$ and with the
recursion equation for $k>1$,
\[
\conn^{k}=\frac{1}{k-1}\sum_{i=1}^{k-1}
 (\conn^i\tens \conn^{k-i})\circ \Qop .
\]
\end{prop}
Given the definition of $\conn^k$ in (\ref{eq:conn1pi})
Proposition~\ref{prop:recconn} is implied by the
following lemma.
\begin{lem}
\[
\Dop^{k-1}=\frac{1}{k-1}\sum_{i=1}^{k-1}
 (\Dop^{i-1}\tens \Dop^{k-i-1})\circ Q\quad\forall k>1 .
\]
\end{lem} 
\begin{proof}
As a first step to the proof, we ``commute'' the $\Rop$-operator
contained in $\Qop$ through the tensor product $\Dop^{i-1}\tens
\Dop^{k-i-1}$. To this end, recall the multiplicativity of the
(iterated coproduct) together with the factorization property of
$\Dop$ (Lemma~\ref{lem:lambdafac}). We obtain the equivalent
expression
\begin{align}\nonumber
\Dop^{k-1} & =\frac{1}{2(k-1)}\sum_{i=1}^{k-1}
 \left((\cop^{i-1}\tens\cop^{k-i-1}) \Rop\right)\cdot
(\Dop^{i-1}\tens \Dop^{k-i-1})\circ \cop
\nonumber\\
 & =\frac{1}{2(k-1)}\sum_{i=1}^{k-1}
 \left(\sum_{a=1}^{i}\sum_{b=i+1}^{k} \Rop_{a,b}\right)\cdot
(\Dop^{i-1}\tens \Dop^{k-i-1})\circ \cop .
\label{eq:pairgraph2}
\end{align}
Before proceeding with the proof we note that this formula has a
straightforward interpretation in terms of sums over weighted tree
diagrams following the correspondence of
Section~\ref{sec:conn1pi}. Namely, the formula states that the
weighted sum over trees with $k$ vertices is given by summing over all
ordered pairs of weighted trees with total number of vertices equal
to $k$, connecting them in all possible ways with an edge and
dividing by $2(k-1)$.\footnote{Indeed, it would be possible to base
the proof of Lemma~\ref{lem:weightwlegs} on the recursion formula for
$\Dop$ given here instead of (\ref{eq:reclambda}). The argument would
then roughly proceed by considering all $k-1$ ways to cut a tree with
$k$ vertices into two by removing an internal edge. The factor $2$
accounts for the relative ordering of the two subtrees.}

The equation (\ref{eq:pairgraph2}) is proved by induction. We verify
it for $k=2$,
$$\Dop^1=\frac{1}{2}\Rop\cdot(\Dop^0\otimes\Dop^0)\circ
\cop=\frac{1}{2}\Rop\cdot\cop\,,$$
and assume it holds for general order $k$. Then, (\ref{eq:reclambda})
yields
\begin{flalign*}
& \Dop^{k} = \frac{1}{k}\biggl(\sum_{j=1}^{k}\Qop_j\biggr)\circ\Dop^{k-1}\\
& =\frac{1}{k}\biggl(\sum_{j=1}^{k}\Qop_j\biggr)
\frac{1}{2(k-1)}\left(\sum_{i=1}^{k-1}\biggl(\sum_{a=1}^{i}\sum_{b=i+1}^{k}\,
\Rop_{a,b}\biggr)\cdot\biggr(\Dop^{i-1}\otimes\Dop^{k-i-1}\biggr)\right)
\circ\cop\\
& =  \frac{1}{2k(k-1)}\sum_{i=1}^{k-1}\biggl(\biggl(\sum_{j=1}^{i}\Qop_j\biggr)\biggl(\biggl(\sum_{a=1}^{i}\sum_{b=i+1}^{k}\Rop_{a,b}\biggr)\cdot(\Dop^{i-1}\otimes\Dop^{k-1-i})\biggr) \\
&\quad  + \biggl(\sum_{j=i+1}^{k}\Qop_j\biggr)\biggl(\biggl(\sum_{a=1}^{i}\sum_{b=i+1}^{k}\Rop_{a,b}\biggr)\cdot(\Dop^{i-1}\otimes\Dop^{k-1-i})\biggr)\biggr) \circ\cop\\
& =  \frac{1}{2k(k-1)}\sum_{i=1}^{k-1}\biggl(\sum_{j=1}^{i}\biggl(\biggl(\cop_j\sum_{a=1}^{i}\sum_{b=i+1}^{k}\Rop_{a,b}\biggr)\cdot\Qop_j\biggr)\cdot(\Dop^{i-1}\otimes\Dop^{k-1-i})\\
&\quad  + \sum_{j=i+1}^{k}\biggl(\biggl(\cop_j\sum_{a=1}^{i}\sum_{b=i+1}^{k}\Rop_{a,b}\biggr)\cdot\Qop_j\biggr)\cdot(\Dop^{i-1}\otimes\Dop^{k-1-i})\biggr)\circ\cop\\
& =  \frac{1}{2k(k-1)}\sum_{i=1}^{k-1}\biggl(i\biggl(\sum_{a=1}^{i+1}\sum_{b=i+2}^{k+1}\, \Rop_{a,b}\biggr)\cdot(\Dop^{i}\otimes\Dop^{k-1-i})\\
&\quad  + (k-i)\biggl(\sum_{a=1}^{i}\sum_{b=i+1}^{k+1}\,\Rop_{a,b}\biggr)\cdot(\Dop^{i-1}\otimes\Dop^{k-i})\biggr)\circ\cop\\
& =  \frac{1}{2k}\biggl(\biggl(\sum_{a=1}^{k}\, \Rop_{a,k+1}\biggr)\cdot(\Dop^{k-1}\otimes\Dop^{0})+\sum_{i=2}^{k-1}\biggl(\sum_{a=1}^{i}\sum_{b=i+1}^{k+1}\, \Rop_{a,b}\biggr)\cdot(\Dop^{i-1}\otimes\Dop^{k-i})\\
&\quad  +\biggl(\sum_{b=2}^{k+1}\,\Rop_{1,b}\biggr)\cdot(\Dop^{0}\otimes\Dop^{k-1})\biggr)\circ\cop\\
& = \frac{1}{2k}\sum_{i=1}^{k}\left(\sum_{a=1}^{i}\sum_{b=i+1}^{k+1}\,
\Rop_{a,b}\right)\cdot\left(\Dop^{i-1}\otimes\Dop^{k-i}\right)\circ\cop
.
\end{flalign*}
\end{proof}

\section{Extensions and applications}
\label{sec:discuss}

\subsection{Modified 1PI functions}

In Section~\ref{sec:review} we have discussed two versions of 1PI
functions, the standard one, denoted $\Gopi$, and a modified one,
denoted $\Gopim$.
Recall that the connected $n$-point functions $\Gconn$ are expressible
in terms of $\Gopi$ and of $\Gopim$ essentially in the same way, as a
sum over all tree graphs, the difference being that for $\Gopim$ the
(inverse) Feynman propagator $\Gfeyn$ is replaced by the (inverse)
connected propagator $\Gconn^{(2)}$ for all edges. Thus, the results of
Sections~\ref{sec:conn1pi} and \ref{sec:recrel} immediately carry over
to the relation between connected functions and modified 1PI functions.
We only need to modify the
definition of $\Rop$ by replacing $\Gfeyn^{-1}$ with
$\Gconn^{(2)\,-1}$ in (\ref{eq:R}); call the new version $\Ropm$. We
denote the induced modifications of $\Qop$ and $\Dop$ by $\Qopm$ and
$\Dopm$ respectively.
\begin{cor}
\label{cor:conn1pim}
The connected $n$-point functions may be expressed in terms of the
modified 1PI ones through the formula
\begin{equation}
\conn=\sum_{k=1}^\infty \connm^k,\quad\text{with}\quad
 \connm^k\defeq \opim^{\tens k}\circ\Dopm^{k-1} .
\label{eq:conn1pim}
\end{equation}
\end{cor}
\begin{cor}
\label{prop:recconnm}
$\connm^k$ may be determined recursively via $\connm^1=\opim$ and with the
recursion equation for $k>1$,
\[
\connm^{k}=\frac{1}{k-1}\sum_{i=1}^{k-1}
 (\connm^i\tens \connm^{k-i})\circ \Qopm .
\]
\end{cor}

Recall that for the modified 1PI functions not only $0$- and
$1$-point functions vanish, but also $2$-point functions.
This
implies that only trees contribute which have
the property that all their vertices have valence at least
three. For a given number of external legs, there are only finitely
many such trees.
Thus, in contrast to (\ref{eq:conn1pi})
the sum in (\ref{eq:conn1pim}) is finite for each given
set of external legs, i.e., for each element of $\SV$ to which it
is applied.

\subsection{Tree level contributions}

Another context in quantum field theory, where we sum over all
tree diagrams is of course when we wish to evaluate merely the tree
level contribution to an $n$-point function. Let us indicate the tree
level contribution by a lower index ``T''. Thus, the tree level
contributions to the algebraic $n$-point functions are denoted
$\complt$, $\connt$ and $\opit$ for the complete, connected and 1PI
$n$-point functions respectively. Of course, the latter are now
nothing but the interaction terms of the Lagrangian. The results of
Section~\ref{sec:complconn} carry over immediately.
\begin{cor}
$\complt=\exp_\star \connt$, where the convolution exponential is defined
in terms of its power series expansion.
\end{cor}
\begin{cor}
$\connt=\log_\star \complt$, where the convolution logarithm is defined
in terms of its power series expansion in $\complt-\cou$.
\end{cor}

The results of Sections~\ref{sec:conn1pi} and \ref{sec:recrel} take
the following form.
\begin{cor}
\label{cor:conn1pit}
The tree level connected $n$-point functions may be expressed in terms
of the interaction vertices through the formula
\begin{equation}
\connt=\sum_{k=1}^\infty \connt^k,\quad\text{with}\quad
 \connt^k\defeq \opit^{\tens k}\circ\Dop^{k-1} .
\label{eq:conn1pit}
\end{equation}
\end{cor}
\begin{cor}
\label{prop:recconnt}
$\connt^k$ may be determined recursively via $\connt^1=\opit$ and with the
recursion equation for $k>1$,
\[
\connt^{k}=\frac{1}{k-1}\sum_{i=1}^{k-1}
 (\connt^i\tens \connt^{k-i})\circ Q .
\]
\end{cor}
Not only the $0$- and $1$-point, but also the $2$-point contribution to
$\opit$ vanishes by definition. Thus, as in the case of the modified
1PI functions, only finitely many tree graphs with given external legs
contribute and the sum (\ref{eq:conn1pit}) is finite on any given
element of $\SV$.

\subsection{Algorithmic Considerations}
\label{sec:algo}

A key feature of the Hopf algebraic approach to $n$-point
functions presented here is the close relation of the obtained
algebraic relations to concrete algorithms. In
Section~\ref{sec:conn1pi} the recursive definition
(\ref{eq:reclambda}) mirrors an algorithm to construct all tree
graphs.
Tree graphs with $k+1$ vertices are created from those
with $k$ vertices by taking each graph in turn and applying the
following procedure: Take every vertex of the graph in turn and split
it into two vertices, distributing the legs in all possible ways and
reconnecting the two new vertices.

Of course, the actual algorithm is slightly more complicated as
symmetries have to be taken into account and the correct weights must
be obtained. However, the recursion relation (\ref{eq:reclambda}) even
suggests an implementation of data structures. For example, we
may represent $\SV^{\tens n}$ by an array, each element of which
corresponds to a graph in the sense of Section~\ref{sec:conn1pi}. Such
an element would contain a rational scalar (the weight) and an array,
each element of which would correspond to one tensor factor in
$\SV^{\tens n}$ or equivalently to one vertex. In turn each element
would be a set of symbolic elements representing external or internal
legs. The coproduct and the $\Rop$-operator are then very simple
operations distributing or adding pairs of symbolic elements.

In this context a certain property of the recursion algorithm
represented by (\ref{eq:reclambda}) is rather interesting. Namely, it
is easy to see that a graph containing at least one vertex with
valence $1$ will in a recursion step
only generate graphs that contain at least one vertex
with valence $1$. Conversely, this means that if we are not
interested in such graphs we may exclude them at each
recursion step without losing any relevant graphs. What is more, we
may implement this removal of ``irrelevant'' graphs through a
\emph{truncated coproduct}. Recall from (\ref{eq:coproduct}) that the
coproduct restricted to the subspace $V^n\subset\SV$ is a map
$V^n\to \bigoplus_{i=0}^n V^i\tens V^{n-i}$. Removing those components
of the direct sum where at least one of the target tensor factors is
$V^0$ we obtain a map $V^n\to \bigoplus_{i=1}^{n-1} V^i\tens
V^{n-i}$. We call this the truncated coproduct $\cop_{\ge 1}$. For
example,
\begin{gather*}
\cop_{\ge 1}(\one)=0,\qquad \cop_{\ge 1}(\phi(x))=0,\\
\cop_{\ge 1}(\phi(x)\phi(y))=\phi(x)\tens\phi(y)
 +\phi(y)\tens\phi(x)
\end{gather*}
In the recursion process (\ref{eq:reclambda}) the number of legs of a
vertex changes when the $\Qop$-operator is applied to it. The only
terms corresponding to $1$-point vertices arise from those terms in
the coproduct where one of the new vertices receives no legs at all
and has thus, after reconnecting with $\Rop$, only one leg. Thus,
replacing the coproduct $\cop$ by the truncated coproduct $\cop_{\ge
1}$ exactly eliminates the irrelevant trees with $1$-point
vertices. 

If we limit the allowed valence of vertices even more, we can push the
truncation prescription even further. Assume we are interested only in
trees with vertices of valence at least three (as in the case of
modified 1PI functions or tree level calculations). By extension of
the above discussion it is clear that the removal of all irrelevant
trees is achieved by a further truncation of the coproduct. Namely,
remove from the coproduct map $V^n\to \bigoplus_{i=0}^n V^i\tens
V^{n-i}$ the components with tensor factors $V^0$ and $V^1$. We denote
the truncated coproduct defined in this way by $\cop_{\ge 2}$. It is
then obvious that using this truncated coproduct to define $\Qop$ and
in turn $\Dop$ produces only trees all of whose vertices have valence
at least three. In particular, this means that for given external legs
the algorithm sketched above for calculating all trees terminates
after finitely many steps. As in the above case, it never creates a
tree that would need to be discarded later. Note that this procedure
may be extended to any lower bound $n$ on the valence of vertices by
using the corresponding $(n-1)$-truncated coproduct $\cop_{\ge n-1}$
defined in the obvious way.

\subsection{Fermions}
\label{sec:fermions}

Recall that we have limited ourselves above to a purely bosonic
theory. However, as already mentioned, this limitation is purely one
of convenience and simplicity. Indeed, all of our arguments and
results apply equally to fermionic fields. However, the underlying
formalism becomes slightly more complicated. The vector space $V$ will
in general be a \emph{$\Z_2$-graded} vector space, a direct sum of a
bosonic and a fermionic part. In turn, the algebra $\SV$ is the
\emph{$\Z_2$-graded}
symmetric algebra over $V$. As special cases, if $V$ is completely
bosonic we recover the usual commutative symmetric algebra (as above);
if $V$ is completely fermionic we recover the usual anti-commutative
exterior algebra.
As a Hopf algebra $\SV$ is in general a \emph{$\Z_2$-graded} or
\emph{super-}Hopf algebra. In particular, the coproduct becomes
graded. This simply means that a minus sign appears as soon as odd
elements are commuted, e.g.,
\[
\cop (ab)=ab\tens \one + \one\tens ab + a\tens b
 + (-1)^{|a| |b|} b\tens a .
\]
Here, $|a|$ is defined to be $0$ or $1$ depending on whether $a$ is
bosonic or fermionic.
We refer to \cite{BFFO:twist} for more details on
the structure of $\SV$ in general.

However, all formulas appearing in Theorems,
Propositions and Corollaries generalize completely unchanged. The
graded structure is completely implicit there. The only explicitly
changing formulas are indeed those that involve explicit evaluations
of the coproduct such as (\ref{eq:coptwo}), (\ref{eq:coproduct}) and
(\ref{eq:copk}). The underlying reason is that our constructions are
completely ``functorial'' and could indeed be generalized to arbitrary
(reasonable) symmetric categories. (For a generalization of certain
$n$-point functions to non-symmetric categories see \cite{Oe:bqft}.)

\section{Conclusions}
\label{sec:concl}

Functional methods used to handle the combinatorics of quantum field
theory, while having a certain elegance, have some serious
drawbacks. While appearing to be analytic, they are really formal as
the mathematical objects involved usually do not actually
exist. In particular, the source terms appearing in functional
expressions are usually merely a book keeping device, rather than
actual mathematical (or physical) entities.
We hope to have convinced the reader that at least certain
combinatorial aspects of quantum field theory can be handled in a
much more intrinsic and (we think) at least as elegant
language. Indeed, our main object is nothing but the rather concrete
time-ordered algebra of field operators. Its coproduct, while perhaps
an unusual structure for quantum field theorists, is well known to
mathematicians. It is thus natural to use it instead of more indirect
and formal functional methods.

Another advantage of our algebraic approach over a functional one is
its closeness to algorithmic descriptions of the processes
involved. Recall from Section~\ref{sec:algo} how easy it is to
translate the recursion relation underlying the correspondence between
connected and 1PI $n$-point functions into an algorithm, which
moreover appears to be rather efficient.

As mentioned in the introduction, the present paper shares a common
programme with \cite{BFFO:twist}, namely to employ the full Hopf
algebraic structure of the algebra of field operators in describing
and understanding quantum field theory. Thus, it is natural to combine
the results of the present paper with those of
\cite{BFFO:twist}. Indeed, the functorial nature of the Drinfeld twist
employed in \cite{BFFO:twist} to relate different products and their
(complete) $n$-point functions should make it possible to induce the
corresponding transformation on the corresponding connected or 1PI
$n$-point functions using the results presented here. This, of course,
goes beyond the scope of the present paper.

\subsection*{Acknowledgements}

\^A.~M.\ was supported through fellowships provided by Funda\c c\~ao
para a Ci\^encia e a Tecnologia SFRH/BD/1282/2000 and subsequently
by Funda\c c\~ao Calouste Gulbenkian 65709.

\appendix
\section{Appendix}

This appendix shows all tree graphs without external legs and
with up to $7$ vertices, computed as contributions to $\Dop^k(\one)$
via (\ref{eq:reclambda}). The factors in front are the inverses of the
symmetry factors of Definition~\ref{def:symfac}, see
Lemma~\ref{lem:weightgeneral}.
\vspace{1cm}\\
\setlength{\unitlength}{0.00083333in}
\begingroup\makeatletter\ifx\SetFigFont\undefined%
\gdef\SetFigFont#1#2#3#4#5{%
  \reset@font\fontsize{#1}{#2pt}%
  \fontfamily{#3}\fontseries{#4}\fontshape{#5}%
  \selectfont}%
\fi\endgroup%
{\renewcommand{\dashlinestretch}{30}
\begin{picture}(1066,261)(0,-10)
\texture{44555555 55aaaaaa aa555555 55aaaaaa aa555555 55aaaaaa aa555555 55aaaaaa 
	aa555555 55aaaaaa aa555555 55aaaaaa aa555555 55aaaaaa aa555555 55aaaaaa 
	aa555555 55aaaaaa aa555555 55aaaaaa aa555555 55aaaaaa aa555555 55aaaaaa 
	aa555555 55aaaaaa aa555555 55aaaaaa aa555555 55aaaaaa aa555555 55aaaaaa }
\put(983,163){\shade\ellipse{150}{150}}
\put(983,163){\ellipse{150}{150}}
\put(0,58){\makebox(0,0)[lb]{{\SetFigFont{12}{14.4}{\rmdefault}{\mddefault}{\updefault}$\Dop^0(\one)\,\,=$}}}
\end{picture}
}
\vspace{1cm}\\
\setlength{\unitlength}{0.00083333in}
\begingroup\makeatletter\ifx\SetFigFont\undefined%
\gdef\SetFigFont#1#2#3#4#5{%
  \reset@font\fontsize{#1}{#2pt}%
  \fontfamily{#3}\fontseries{#4}\fontshape{#5}%
  \selectfont}%
\fi\endgroup%
{\renewcommand{\dashlinestretch}{30}
\begin{picture}(1545,269)(0,-10)
\texture{44555555 55aaaaaa aa555555 55aaaaaa aa555555 55aaaaaa aa555555 55aaaaaa 
	aa555555 55aaaaaa aa555555 55aaaaaa aa555555 55aaaaaa aa555555 55aaaaaa 
	aa555555 55aaaaaa aa555555 55aaaaaa aa555555 55aaaaaa aa555555 55aaaaaa 
	aa555555 55aaaaaa aa555555 55aaaaaa aa555555 55aaaaaa aa555555 55aaaaaa }
\path(1177,163)(1477,163)
\path(1177,163)(1477,163)
\put(1162,171){\shade\ellipse{150}{150}}
\put(1162,171){\ellipse{150}{150}}
\put(1462,171){\shade\ellipse{150}{150}}
\put(1462,171){\ellipse{150}{150}}
\put(0,58){\makebox(0,0)[lb]{{\SetFigFont{12}{14.4}{\rmdefault}{\mddefault}{\updefault}$\Dop^1(\one)\,\,=\,\,\frac{1}{2}$}}}
\end{picture}
}
\vspace{1cm}\\
\setlength{\unitlength}{0.00083333in}
\begingroup\makeatletter\ifx\SetFigFont\undefined%
\gdef\SetFigFont#1#2#3#4#5{%
  \reset@font\fontsize{#1}{#2pt}%
  \fontfamily{#3}\fontseries{#4}\fontshape{#5}%
  \selectfont}%
\fi\endgroup%
{\renewcommand{\dashlinestretch}{30}
\begin{picture}(1845,269)(0,-10)
\texture{44555555 55aaaaaa aa555555 55aaaaaa aa555555 55aaaaaa aa555555 55aaaaaa 
	aa555555 55aaaaaa aa555555 55aaaaaa aa555555 55aaaaaa aa555555 55aaaaaa 
	aa555555 55aaaaaa aa555555 55aaaaaa aa555555 55aaaaaa aa555555 55aaaaaa 
	aa555555 55aaaaaa aa555555 55aaaaaa aa555555 55aaaaaa aa555555 55aaaaaa }
\path(1162,171)(1462,171)
\path(1162,171)(1462,171)
\path(1462,171)(1762,171)
\path(1462,171)(1762,171)
\put(1162,171){\shade\ellipse{150}{150}}
\put(1162,171){\ellipse{150}{150}}
\put(1462,171){\shade\ellipse{150}{150}}
\put(1462,171){\ellipse{150}{150}}
\put(1762,171){\shade\ellipse{150}{150}}
\put(1762,171){\ellipse{150}{150}}
\put(0,58){\makebox(0,0)[lb]{{\SetFigFont{12}{14.4}{\rmdefault}{\mddefault}{\updefault}$\Dop^2(\one)\,\,=\,\,\frac{1}{2}$}}}
\end{picture}
}
\vspace{1cm}\\
\setlength{\unitlength}{0.00083333in}
\begingroup\makeatletter\ifx\SetFigFont\undefined%
\gdef\SetFigFont#1#2#3#4#5{%
  \reset@font\fontsize{#1}{#2pt}%
  \fontfamily{#3}\fontseries{#4}\fontshape{#5}%
  \selectfont}%
\fi\endgroup%
{\renewcommand{\dashlinestretch}{30}
\begin{picture}(3353,701)(0,-10)
\texture{44555555 55aaaaaa aa555555 55aaaaaa aa555555 55aaaaaa aa555555 55aaaaaa 
	aa555555 55aaaaaa aa555555 55aaaaaa aa555555 55aaaaaa aa555555 55aaaaaa 
	aa555555 55aaaaaa aa555555 55aaaaaa aa555555 55aaaaaa aa555555 55aaaaaa 
	aa555555 55aaaaaa aa555555 55aaaaaa aa555555 55aaaaaa aa555555 55aaaaaa }
\path(3270,343)(2970,343)
\path(3270,343)(2970,343)
\path(2820,603)(2980,333)
\path(2830,93)(2990,363)
\path(1207,335)(1507,335)
\path(1207,335)(1507,335)
\path(1507,335)(1807,335)
\path(1507,335)(1807,335)
\path(1807,335)(2107,335)
\path(1807,335)(2107,335)
\put(2970,343){\shade\ellipse{150}{150}}
\put(2970,343){\ellipse{150}{150}}
\put(3270,343){\shade\ellipse{150}{150}}
\put(3270,343){\ellipse{150}{150}}
\put(2825,603){\shade\ellipse{150}{150}}
\put(2825,603){\ellipse{150}{150}}
\put(2825,83){\shade\ellipse{150}{150}}
\put(2825,83){\ellipse{150}{150}}
\put(1477,328){\shade\ellipse{150}{150}}
\put(1477,328){\ellipse{150}{150}}
\put(1777,328){\shade\ellipse{150}{150}}
\put(1777,328){\ellipse{150}{150}}
\put(2077,328){\shade\ellipse{150}{150}}
\put(2077,328){\ellipse{150}{150}}
\put(1177,336){\shade\ellipse{150}{150}}
\put(1177,336){\ellipse{150}{150}}
\put(2250,223){\makebox(0,0)[lb]{{\SetFigFont{12}{14.4}{\rmdefault}{\mddefault}{\updefault}$+\,\,\frac{1}{3!}$}}}
\put(0,223){\makebox(0,0)[lb]{{\SetFigFont{12}{14.4}{\rmdefault}{\mddefault}{\updefault}$\Dop^3(\one)\,\,=\,\,\frac{1}{2}$}}}
\end{picture}
}
\vspace{1cm}\\
\setlength{\unitlength}{0.00083333in}
\begingroup\makeatletter\ifx\SetFigFont\undefined%
\gdef\SetFigFont#1#2#3#4#5{%
  \reset@font\fontsize{#1}{#2pt}%
  \fontfamily{#3}\fontseries{#4}\fontshape{#5}%
  \selectfont}%
\fi\endgroup%
{\renewcommand{\dashlinestretch}{30}
\begin{picture}(5306,781)(0,-10)
\texture{44555555 55aaaaaa aa555555 55aaaaaa aa555555 55aaaaaa aa555555 55aaaaaa 
	aa555555 55aaaaaa aa555555 55aaaaaa aa555555 55aaaaaa aa555555 55aaaaaa 
	aa555555 55aaaaaa aa555555 55aaaaaa aa555555 55aaaaaa aa555555 55aaaaaa 
	aa555555 55aaaaaa aa555555 55aaaaaa aa555555 55aaaaaa aa555555 55aaaaaa }
\path(1177,391)(1477,391)
\path(1177,391)(1477,391)
\path(1477,391)(1777,391)
\path(1477,391)(1777,391)
\path(1777,391)(2077,391)
\path(1777,391)(2077,391)
\path(2077,391)(2377,391)
\path(2077,391)(2377,391)
\path(3570,383)(3270,383)
\path(3570,383)(3270,383)
\path(3120,643)(3280,373)
\path(3130,133)(3290,403)
\path(3840,383)(3540,383)
\path(3840,383)(3540,383)
\path(4923,83)(4923,383)
\path(4923,83)(4923,383)
\path(4923,383)(4923,683)
\path(4923,383)(4923,683)
\path(4998,383)(4698,383)
\path(4998,383)(4698,383)
\path(4923,83)(4923,383)
\path(4923,83)(4923,383)
\path(5223,383)(4923,383)
\path(5223,383)(4923,383)
\put(1177,391){\shade\ellipse{150}{150}}
\put(1177,391){\ellipse{150}{150}}
\put(1477,391){\shade\ellipse{150}{150}}
\put(1477,391){\ellipse{150}{150}}
\put(1777,391){\shade\ellipse{150}{150}}
\put(1777,391){\ellipse{150}{150}}
\put(2077,391){\shade\ellipse{150}{150}}
\put(2077,391){\ellipse{150}{150}}
\put(2377,391){\shade\ellipse{150}{150}}
\put(2377,391){\ellipse{150}{150}}
\put(3270,383){\shade\ellipse{150}{150}}
\put(3270,383){\ellipse{150}{150}}
\put(3570,383){\shade\ellipse{150}{150}}
\put(3570,383){\ellipse{150}{150}}
\put(3870,383){\shade\ellipse{150}{150}}
\put(3870,383){\ellipse{150}{150}}
\put(3125,643){\shade\ellipse{150}{150}}
\put(3125,643){\ellipse{150}{150}}
\put(3125,123){\shade\ellipse{150}{150}}
\put(3125,123){\ellipse{150}{150}}
\put(4623,383){\shade\ellipse{150}{150}}
\put(4623,383){\ellipse{150}{150}}
\put(4923,83){\shade\ellipse{150}{150}}
\put(4923,83){\ellipse{150}{150}}
\put(4923,383){\shade\ellipse{150}{150}}
\put(4923,383){\ellipse{150}{150}}
\put(4923,683){\shade\ellipse{150}{150}}
\put(4923,683){\ellipse{150}{150}}
\put(5223,383){\shade\ellipse{150}{150}}
\put(5223,383){\ellipse{150}{150}}
\put(0,270){\makebox(0,0)[lb]{{\SetFigFont{12}{14.4}{\rmdefault}{\mddefault}{\updefault}$\Dop^4(\one)\,\,=\,\,\frac{1}{2}$}}}
\put(2550,263){\makebox(0,0)[lb]{{\SetFigFont{12}{14.4}{\rmdefault}{\mddefault}{\updefault}$+\,\,\frac{1}{2}$}}}
\put(4065,263){\makebox(0,0)[lb]{{\SetFigFont{12}{14.4}{\rmdefault}{\mddefault}{\updefault}$+\,\,\frac{1}{4!}$}}}
\end{picture}
}
\vspace{1cm}\\
\setlength{\unitlength}{0.00083333in}
\begingroup\makeatletter\ifx\SetFigFont\undefined%
\gdef\SetFigFont#1#2#3#4#5{%
  \reset@font\fontsize{#1}{#2pt}%
  \fontfamily{#3}\fontseries{#4}\fontshape{#5}%
  \selectfont}%
\fi\endgroup%
{\renewcommand{\dashlinestretch}{30}
\begin{picture}(5903,2367)(0,-10)
\path(1185,1756)(2685,1756)
\path(1485,383)(2370,383)
\texture{44555555 55aaaaaa aa555555 55aaaaaa aa555555 55aaaaaa aa555555 55aaaaaa 
	aa555555 55aaaaaa aa555555 55aaaaaa aa555555 55aaaaaa aa555555 55aaaaaa 
	aa555555 55aaaaaa aa555555 55aaaaaa aa555555 55aaaaaa aa555555 55aaaaaa 
	aa555555 55aaaaaa aa555555 55aaaaaa aa555555 55aaaaaa aa555555 55aaaaaa }
\path(3862,1763)(3562,1763)
\path(3862,1763)(3562,1763)
\path(3412,2023)(3572,1753)
\path(3422,1513)(3582,1783)
\path(4132,1763)(3832,1763)
\path(4132,1763)(3832,1763)
\path(4462,1758)(4162,1758)
\path(4462,1758)(4162,1758)
\path(3127,658)(3287,388)
\path(3137,148)(3297,418)
\path(3292,391)(3547,391)(3592,406)
\path(3715,140)(3555,410)
\path(3705,650)(3545,380)
\path(4734,381)(5034,381)
\path(4734,381)(5034,381)
\path(4829,671)(4729,341)
\path(4829,91)(4739,391)
\path(4729,391)(4489,191)
\path(4729,371)(4489,551)
\path(5520,1748)(5220,1223)
\path(5380,1498)(5540,1768)
\path(5820,1748)(5520,1748)
\path(5820,1748)(5520,1748)
\path(5370,2008)(5530,1738)
\path(5220,2273)(5520,1748)
\path(1770,83)(1770,383)
\path(1770,83)(1770,383)
\path(1770,383)(1770,683)
\path(1770,383)(1770,683)
\path(1770,83)(1770,383)
\path(1770,83)(1770,383)
\put(3562,1763){\shade\ellipse{150}{150}}
\put(3562,1763){\ellipse{150}{150}}
\put(3862,1763){\shade\ellipse{150}{150}}
\put(3862,1763){\ellipse{150}{150}}
\put(4162,1763){\shade\ellipse{150}{150}}
\put(4162,1763){\ellipse{150}{150}}
\put(3417,2023){\shade\ellipse{150}{150}}
\put(3417,2023){\ellipse{150}{150}}
\put(3417,1503){\shade\ellipse{150}{150}}
\put(3417,1503){\ellipse{150}{150}}
\put(4467,1758){\shade\ellipse{150}{150}}
\put(4467,1758){\ellipse{150}{150}}
\put(3277,398){\shade\ellipse{150}{150}}
\put(3277,398){\ellipse{150}{150}}
\put(3132,658){\shade\ellipse{150}{150}}
\put(3132,658){\ellipse{150}{150}}
\put(3132,138){\shade\ellipse{150}{150}}
\put(3132,138){\ellipse{150}{150}}
\put(3565,400){\shade\ellipse{150}{150}}
\put(3565,400){\ellipse{150}{150}}
\put(3710,140){\shade\ellipse{150}{150}}
\put(3710,140){\ellipse{150}{150}}
\put(3710,660){\shade\ellipse{150}{150}}
\put(3710,660){\ellipse{150}{150}}
\put(4734,381){\shade\ellipse{150}{150}}
\put(4734,381){\ellipse{150}{150}}
\put(5034,381){\shade\ellipse{150}{150}}
\put(5034,381){\ellipse{150}{150}}
\put(4829,86){\shade\ellipse{150}{150}}
\put(4829,86){\ellipse{150}{150}}
\put(4499,186){\shade\ellipse{150}{150}}
\put(4499,186){\ellipse{150}{150}}
\put(4489,556){\shade\ellipse{150}{150}}
\put(4489,556){\ellipse{150}{150}}
\put(4829,666){\shade\ellipse{150}{150}}
\put(4829,666){\ellipse{150}{150}}
\put(5375,1488){\shade\ellipse{150}{150}}
\put(5375,1488){\ellipse{150}{150}}
\put(5229,1231){\shade\ellipse{150}{150}}
\put(5229,1231){\ellipse{150}{150}}
\put(5520,1748){\shade\ellipse{150}{150}}
\put(5520,1748){\ellipse{150}{150}}
\put(5820,1748){\shade\ellipse{150}{150}}
\put(5820,1748){\ellipse{150}{150}}
\put(5375,2008){\shade\ellipse{150}{150}}
\put(5375,2008){\ellipse{150}{150}}
\put(5225,2269){\shade\ellipse{150}{150}}
\put(5225,2269){\ellipse{150}{150}}
\put(1470,383){\shade\ellipse{150}{150}}
\put(1470,383){\ellipse{150}{150}}
\put(1770,83){\shade\ellipse{150}{150}}
\put(1770,83){\ellipse{150}{150}}
\put(1770,683){\shade\ellipse{150}{150}}
\put(1770,683){\ellipse{150}{150}}
\put(2070,383){\shade\ellipse{150}{150}}
\put(2070,383){\ellipse{150}{150}}
\put(2355,376){\shade\ellipse{150}{150}}
\put(2355,376){\ellipse{150}{150}}
\put(1177,1755){\shade\ellipse{150}{150}}
\put(1177,1755){\ellipse{150}{150}}
\put(1477,1755){\shade\ellipse{150}{150}}
\put(1477,1755){\ellipse{150}{150}}
\put(1777,1755){\shade\ellipse{150}{150}}
\put(1777,1755){\ellipse{150}{150}}
\put(2377,1755){\shade\ellipse{150}{150}}
\put(2377,1755){\ellipse{150}{150}}
\put(2677,1755){\shade\ellipse{150}{150}}
\put(2677,1755){\ellipse{150}{150}}
\put(2077,1755){\shade\ellipse{150}{150}}
\put(2077,1755){\ellipse{150}{150}}
\put(1762,383){\shade\ellipse{150}{150}}
\put(1762,383){\ellipse{150}{150}}
\put(2857,1628){\makebox(0,0)[lb]{{\SetFigFont{12}{14.4}{\rmdefault}{\mddefault}{\updefault}$+\,\,\frac{1}{2}$}}}
\put(907,271){\makebox(0,0)[lb]{{\SetFigFont{12}{14.4}{\rmdefault}{\mddefault}{\updefault}$+\,\,\frac{1}{3!}$}}}
\put(3915,271){\makebox(0,0)[lb]{{\SetFigFont{12}{14.4}{\rmdefault}{\mddefault}{\updefault}$+\,\,\frac{1}{5!}$}}}
\put(0,1621){\makebox(0,0)[lb]{{\SetFigFont{12}{14.4}{\rmdefault}{\mddefault}{\updefault}$\Dop^5(\one)\,\,=\,\,\frac{1}{2}$}}}
\put(2565,271){\makebox(0,0)[lb]{{\SetFigFont{12}{14.4}{\rmdefault}{\mddefault}{\updefault}$+\,\,\frac{1}{2^3}$}}}
\put(4657,1621){\makebox(0,0)[lb]{{\SetFigFont{12}{14.4}{\rmdefault}{\mddefault}{\updefault}$+\,\,\frac{1}{2}$}}}
\end{picture}
}
\vspace{1cm}\\
\setlength{\unitlength}{0.00083333in}
\begingroup\makeatletter\ifx\SetFigFont\undefined%
\gdef\SetFigFont#1#2#3#4#5{%
  \reset@font\fontsize{#1}{#2pt}%
  \fontfamily{#3}\fontseries{#4}\fontshape{#5}%
  \selectfont}%
\fi\endgroup%
{\renewcommand{\dashlinestretch}{30}
\begin{picture}(5891,5103)(0,-10)
\path(1605,380)(2235,380)
\path(1290,3223)(2190,3223)
\path(1440,1715)(2610,1715)
\path(3232,395)(3817,395)
\path(4560,3245)(5730,3245)
\path(3855,4738)(5085,4738)
\path(4912,3530)(4912,2945)
\texture{44555555 55aaaaaa aa555555 55aaaaaa aa555555 55aaaaaa aa555555 55aaaaaa 
	aa555555 55aaaaaa aa555555 55aaaaaa aa555555 55aaaaaa aa555555 55aaaaaa 
	aa555555 55aaaaaa aa555555 55aaaaaa aa555555 55aaaaaa aa555555 55aaaaaa 
	aa555555 55aaaaaa aa555555 55aaaaaa aa555555 55aaaaaa aa555555 55aaaaaa }
\path(3862,1738)(3562,1738)
\path(3862,1738)(3562,1738)
\path(3412,1998)(3572,1728)
\path(3422,1488)(3582,1758)
\path(4162,1731)(3862,1731)
\path(4162,1731)(3862,1731)
\path(4332,1468)(4172,1738)
\path(4322,1978)(4162,1708)
\path(5212,1998)(5372,1728)
\path(5222,1488)(5382,1758)
\path(5377,1731)(5632,1731)(5677,1746)
\path(5362,1748)(5062,2258)(5052,2258)
\path(5800,1480)(5640,1750)
\path(5790,1990)(5630,1720)
\path(4619,385)(4919,385)
\path(4619,385)(4919,385)
\path(4919,385)(5219,385)
\path(4919,385)(5219,385)
\path(5069,630)(4919,380)
\path(4939,380)(5079,120)
\path(4909,400)(4769,130)
\path(4769,640)(4939,360)
\path(1154,4737)(1454,4737)
\path(1154,4737)(1454,4737)
\path(1454,4737)(1754,4737)
\path(1454,4737)(1754,4737)
\path(1754,4737)(2054,4737)
\path(1754,4737)(2054,4737)
\path(2054,4737)(2354,4737)
\path(2054,4737)(2354,4737)
\path(2354,4737)(2654,4737)
\path(2354,4737)(2654,4737)
\path(2654,4737)(2954,4737)
\path(2654,4737)(2954,4737)
\path(3555,3238)(3255,3238)
\path(3555,3238)(3255,3238)
\path(3105,3498)(3265,3228)
\path(2955,3763)(3255,3238)
\path(3255,3238)(2955,2713)
\path(3115,2988)(3275,3258)
\path(3855,3238)(3555,3238)
\path(3855,3238)(3555,3238)
\path(3712,5005)(3872,4735)
\path(3722,4495)(3882,4765)
\path(2065,1423)(2065,1723)
\path(2065,1423)(2065,1723)
\path(2065,1723)(2065,2023)
\path(2065,1723)(2065,2023)
\path(2065,1423)(2065,1723)
\path(2065,1423)(2065,1723)
\path(1462,648)(1622,378)
\path(1472,138)(1632,408)
\path(1912,663)(1912,103)
\drawline(1912,103)(1912,103)
\path(1155,3491)(1315,3221)
\path(1005,3756)(1305,3231)
\path(1165,2981)(1325,3251)
\path(3324,695)(3224,365)
\path(3324,115)(3234,415)
\path(3224,415)(2984,215)
\path(3224,395)(2984,575)
\put(3562,1738){\shade\ellipse{150}{150}}
\put(3562,1738){\ellipse{150}{150}}
\put(3862,1738){\shade\ellipse{150}{150}}
\put(3862,1738){\ellipse{150}{150}}
\put(3417,1998){\shade\ellipse{150}{150}}
\put(3417,1998){\ellipse{150}{150}}
\put(3417,1478){\shade\ellipse{150}{150}}
\put(3417,1478){\ellipse{150}{150}}
\put(4169,1741){\shade\ellipse{150}{150}}
\put(4169,1741){\ellipse{150}{150}}
\put(4327,1468){\shade\ellipse{150}{150}}
\put(4327,1468){\ellipse{150}{150}}
\put(4327,1988){\shade\ellipse{150}{150}}
\put(4327,1988){\ellipse{150}{150}}
\put(5362,1738){\shade\ellipse{150}{150}}
\put(5362,1738){\ellipse{150}{150}}
\put(5217,1998){\shade\ellipse{150}{150}}
\put(5217,1998){\ellipse{150}{150}}
\put(5217,1478){\shade\ellipse{150}{150}}
\put(5217,1478){\ellipse{150}{150}}
\put(5057,2258){\shade\ellipse{150}{150}}
\put(5057,2258){\ellipse{150}{150}}
\put(5650,1740){\shade\ellipse{150}{150}}
\put(5650,1740){\ellipse{150}{150}}
\put(5795,1480){\shade\ellipse{150}{150}}
\put(5795,1480){\ellipse{150}{150}}
\put(5795,2000){\shade\ellipse{150}{150}}
\put(5795,2000){\ellipse{150}{150}}
\put(4619,385){\shade\ellipse{150}{150}}
\put(4619,385){\ellipse{150}{150}}
\put(4919,385){\shade\ellipse{150}{150}}
\put(4919,385){\ellipse{150}{150}}
\put(5219,385){\shade\ellipse{150}{150}}
\put(5219,385){\ellipse{150}{150}}
\put(4769,635){\shade\ellipse{150}{150}}
\put(4769,635){\ellipse{150}{150}}
\put(5059,635){\shade\ellipse{150}{150}}
\put(5059,635){\ellipse{150}{150}}
\put(5089,135){\shade\ellipse{150}{150}}
\put(5089,135){\ellipse{150}{150}}
\put(4769,125){\shade\ellipse{150}{150}}
\put(4769,125){\ellipse{150}{150}}
\put(1154,4737){\shade\ellipse{150}{150}}
\put(1154,4737){\ellipse{150}{150}}
\put(1454,4737){\shade\ellipse{150}{150}}
\put(1454,4737){\ellipse{150}{150}}
\put(1754,4737){\shade\ellipse{150}{150}}
\put(1754,4737){\ellipse{150}{150}}
\put(2054,4737){\shade\ellipse{150}{150}}
\put(2054,4737){\ellipse{150}{150}}
\put(2354,4737){\shade\ellipse{150}{150}}
\put(2354,4737){\ellipse{150}{150}}
\put(2654,4737){\shade\ellipse{150}{150}}
\put(2654,4737){\ellipse{150}{150}}
\put(2954,4737){\shade\ellipse{150}{150}}
\put(2954,4737){\ellipse{150}{150}}
\put(3255,3238){\shade\ellipse{150}{150}}
\put(3255,3238){\ellipse{150}{150}}
\put(3555,3238){\shade\ellipse{150}{150}}
\put(3555,3238){\ellipse{150}{150}}
\put(3110,3498){\shade\ellipse{150}{150}}
\put(3110,3498){\ellipse{150}{150}}
\put(2960,3759){\shade\ellipse{150}{150}}
\put(2960,3759){\ellipse{150}{150}}
\put(2964,2721){\shade\ellipse{150}{150}}
\put(2964,2721){\ellipse{150}{150}}
\put(3110,2978){\shade\ellipse{150}{150}}
\put(3110,2978){\ellipse{150}{150}}
\put(3855,3238){\shade\ellipse{150}{150}}
\put(3855,3238){\ellipse{150}{150}}
\put(4608,3253){\shade\ellipse{150}{150}}
\put(4608,3253){\ellipse{150}{150}}
\put(4908,2953){\shade\ellipse{150}{150}}
\put(4908,2953){\ellipse{150}{150}}
\put(4908,3253){\shade\ellipse{150}{150}}
\put(4908,3253){\ellipse{150}{150}}
\put(4908,3553){\shade\ellipse{150}{150}}
\put(4908,3553){\ellipse{150}{150}}
\put(5208,3253){\shade\ellipse{150}{150}}
\put(5208,3253){\ellipse{150}{150}}
\put(5493,3246){\shade\ellipse{150}{150}}
\put(5493,3246){\ellipse{150}{150}}
\put(5808,3246){\shade\ellipse{150}{150}}
\put(5808,3246){\ellipse{150}{150}}
\put(3862,4745){\shade\ellipse{150}{150}}
\put(3862,4745){\ellipse{150}{150}}
\put(4162,4745){\shade\ellipse{150}{150}}
\put(4162,4745){\ellipse{150}{150}}
\put(4462,4745){\shade\ellipse{150}{150}}
\put(4462,4745){\ellipse{150}{150}}
\put(3717,5005){\shade\ellipse{150}{150}}
\put(3717,5005){\ellipse{150}{150}}
\put(3717,4485){\shade\ellipse{150}{150}}
\put(3717,4485){\ellipse{150}{150}}
\put(4767,4740){\shade\ellipse{150}{150}}
\put(4767,4740){\ellipse{150}{150}}
\put(5067,4740){\shade\ellipse{150}{150}}
\put(5067,4740){\ellipse{150}{150}}
\put(1765,1723){\shade\ellipse{150}{150}}
\put(1765,1723){\ellipse{150}{150}}
\put(2065,1423){\shade\ellipse{150}{150}}
\put(2065,1423){\ellipse{150}{150}}
\put(2065,1723){\shade\ellipse{150}{150}}
\put(2065,1723){\ellipse{150}{150}}
\put(2065,2023){\shade\ellipse{150}{150}}
\put(2065,2023){\ellipse{150}{150}}
\put(2365,1723){\shade\ellipse{150}{150}}
\put(2365,1723){\ellipse{150}{150}}
\put(2650,1716){\shade\ellipse{150}{150}}
\put(2650,1716){\ellipse{150}{150}}
\put(1465,1731){\shade\ellipse{150}{150}}
\put(1465,1731){\ellipse{150}{150}}
\put(1612,388){\shade\ellipse{150}{150}}
\put(1612,388){\ellipse{150}{150}}
\put(1912,388){\shade\ellipse{150}{150}}
\put(1912,388){\ellipse{150}{150}}
\put(1467,648){\shade\ellipse{150}{150}}
\put(1467,648){\ellipse{150}{150}}
\put(1467,128){\shade\ellipse{150}{150}}
\put(1467,128){\ellipse{150}{150}}
\put(1917,683){\shade\ellipse{150}{150}}
\put(1917,683){\ellipse{150}{150}}
\put(1917,83){\shade\ellipse{150}{150}}
\put(1917,83){\ellipse{150}{150}}
\put(2220,388){\shade\ellipse{150}{150}}
\put(2220,388){\ellipse{150}{150}}
\put(1305,3231){\shade\ellipse{150}{150}}
\put(1305,3231){\ellipse{150}{150}}
\put(1605,3231){\shade\ellipse{150}{150}}
\put(1605,3231){\ellipse{150}{150}}
\put(1160,3491){\shade\ellipse{150}{150}}
\put(1160,3491){\ellipse{150}{150}}
\put(1010,3752){\shade\ellipse{150}{150}}
\put(1010,3752){\ellipse{150}{150}}
\put(1160,2971){\shade\ellipse{150}{150}}
\put(1160,2971){\ellipse{150}{150}}
\put(1905,3224){\shade\ellipse{150}{150}}
\put(1905,3224){\ellipse{150}{150}}
\put(2197,3230){\shade\ellipse{150}{150}}
\put(2197,3230){\ellipse{150}{150}}
\put(3229,405){\shade\ellipse{150}{150}}
\put(3229,405){\ellipse{150}{150}}
\put(3529,405){\shade\ellipse{150}{150}}
\put(3529,405){\ellipse{150}{150}}
\put(3324,110){\shade\ellipse{150}{150}}
\put(3324,110){\ellipse{150}{150}}
\put(2994,210){\shade\ellipse{150}{150}}
\put(2994,210){\ellipse{150}{150}}
\put(2984,580){\shade\ellipse{150}{150}}
\put(2984,580){\ellipse{150}{150}}
\put(3324,690){\shade\ellipse{150}{150}}
\put(3324,690){\ellipse{150}{150}}
\put(3827,402){\shade\ellipse{150}{150}}
\put(3827,402){\ellipse{150}{150}}
\put(0,4618){\makebox(0,0)[lb]{{\SetFigFont{12}{14.4}{\rmdefault}{\mddefault}{\updefault}$\Dop^6(\one)\,\,=\,\,\frac{1}{2}$}}}
\put(3135,4618){\makebox(0,0)[lb]{{\SetFigFont{12}{14.4}{\rmdefault}{\mddefault}{\updefault}$+\,\,\frac{1}{2}$}}}
\put(2385,3118){\makebox(0,0)[lb]{{\SetFigFont{12}{14.4}{\rmdefault}{\mddefault}{\updefault}$+\,\,\frac{1}{3!}$}}}
\put(4050,3118){\makebox(0,0)[lb]{{\SetFigFont{12}{14.4}{\rmdefault}{\mddefault}{\updefault}$+\,\,\frac{1}{3!}$}}}
\put(2842,1610){\makebox(0,0)[lb]{{\SetFigFont{12}{14.4}{\rmdefault}{\mddefault}{\updefault}$+\,\,\frac{1}{2^3}$}}}
\put(4492,1618){\makebox(0,0)[lb]{{\SetFigFont{12}{14.4}{\rmdefault}{\mddefault}{\updefault}$+\,\,\frac{1}{2}$}}}
\put(2392,268){\makebox(0,0)[lb]{{\SetFigFont{12}{14.4}{\rmdefault}{\mddefault}{\updefault}$+\,\,\frac{1}{4!}$}}}
\put(4042,268){\makebox(0,0)[lb]{{\SetFigFont{12}{14.4}{\rmdefault}{\mddefault}{\updefault}$+\,\,\frac{1}{6!}$}}}
\put(885,260){\makebox(0,0)[lb]{{\SetFigFont{12}{14.4}{\rmdefault}{\mddefault}{\updefault}$+\,\,\frac{1}{2\cdot 3!}$}}}
\put(900,1618){\makebox(0,0)[lb]{{\SetFigFont{12}{14.4}{\rmdefault}{\mddefault}{\updefault}$+\,\,\frac{1}{2^2}$}}}
\put(705,3170){\makebox(0,0)[lb]{{\SetFigFont{12}{14.4}{\rmdefault}{\mddefault}{\updefault}$+$}}}
\end{picture}
}
\\

\bibliographystyle{amsordx}
\bibliography{stdrefs}

\end{document}